\documentclass[10pt,aps,prb,amsmath,english,twocolumn,superscriptaddress]{revtex4-1}
\usepackage[T1]{fontenc}
\usepackage[latin1]{inputenc}
\usepackage{graphicx}
\usepackage{amssymb}
\usepackage{dcolumn}
\usepackage{color}
\usepackage{bm}
\usepackage{babel}

\renewcommand{\vec}[1]{\mathrm{\mathbf{#1}}}

\newcommand{\ep}{\epsilon}

\newcommand{\win}{\omega_{L}}
\newcommand{\ein}{\vec{e}_\mathrm{in}}
\newcommand{\eout}{\vec{e}_\mathrm{out}}
\newcommand{\lep}{\lambda\!\!\bar{}\;{}_\ep}

\begin{document}

\title{Raman spectroscopy as a versatile tool for
studying the properties of graphene}
\author{Andrea C. Ferrari}\email{acf26@eng.cam.ac.uk}
\affiliation{Cambridge Graphene Centre, Cambridge University, Cambridge, UK}
\author{Denis M. Basko}
\affiliation{Universit\'e Grenoble 1 and CNRS, LPMMC UMR 5493,
Grenoble, France}

\begin{abstract}
Raman spectroscopy is an integral part of graphene research. It is used to determine the number and orientation of layers, the quality and types of edge, and the effects of perturbations, such as electric and magnetic fields, strain, doping, disorder and functional groups. This, in turn, provides insight into all $sp^2$-bonded carbon allotropes, because graphene is their fundamental building block. Here we review the state of the art, future directions and open questions in Raman spectroscopy of graphene. We describe essential physical processes whose importance has only recently been recognized, such as the various types of resonance at play, and the role of quantum interference. We update all basic concepts and notations, and propose a terminology that is able to describe any result in literature. We finally highlight the potential of Raman spectroscopy for layered materials other than graphene.

\end{abstract}

\maketitle
Graphene is the two-dimensional (2d) building block for $sp^2$ carbon allotropes of every other dimensionality. It can be stacked into 3d graphite, rolled into 1d nanotubes, or wrapped into 0d fullerenes. It is at the centre of an ever expanding research area\cite{Novo2004,GeimRevNM,charlier,bona,bonag}. Near-ballistic transport and high mobility make it an ideal material for nanoelectronics, especially for high frequency applications\cite{Yuming}. Furthermore, its optical and mechanical properties are ideal for micro and nanomechanical systems, thin-film transistors, transparent and conductive composites and electrodes, flexible and printable (opto)electronics, and photonics\cite{GeimRevNM,charlier,bona,torrisi,Sun}.

An ideal characterization tool would be fast and non-destructive, offer high resolution, give structural and electronic information, and be applicable to both laboratory and mass-production scales. Raman spectroscopy\cite{Landsberg,Raman} fulfills all these requirements. The Raman spectrum of graphite was first recorded more than 40 years ago\cite{tk} and, by the time the Raman spectrum of graphene was measured for the first time in 2006\cite{ACFRaman}, Raman spectroscopy had become one of the most popular techniques for the characterization of disordered and amorphous carbons, fullerenes, nanotubes, diamonds, carbon chains, and poly-conjugated molecules\cite{acftrans}. Raman techniques are particularly useful for graphene\cite{Ferrari2007} since the absence of a band gap makes all incident wavelengths resonant, so that the Raman spectrum contains information about both atomic structure and electronic properties. Resonance can also be reached by UV excitation\cite{Ferrari00,Ferrari01}, either with the \textbf{M}-point Van Hove singularity, or in the case of band gap opening, such as in fluorinated-graphene.

The number of graphene layers in a sample can be determined by elastic light scattering/Rayleigh spectroscopy\cite{CasiraghiNL,GeimAPL}, but this approach only works for exfoliated samples on optimized substrates and does not provide other structural or electronic information. Raman spectroscopy, on the other hand, works for all graphene samples\cite{ACFRaman,Ferrari2007}. Moreover, it is able to identify unwanted by-products, structural damage, functional groups and chemical modifications introduced during the preparation, processing or placement of graphene\cite{bonag}. As a result, a Raman spectrum is invaluable for quality control, and for comparing samples used by different research groups.

The toll for the simplicity of Raman measurements is paid when it comes to spectral interpretation. The spectra of all carbon-based materials show only a few prominent features, regardless of the final structure\cite{acftrans}. However, the shapes, intensities and positions of these peaks give a considerable amount of information, often comparable to that obtained by competing techniques that are more complicated and more destructive\cite{acftrans}. For example, Raman spectroscopy can distinguish between a hard amorphous carbon, a metallic nanotube or a doped graphene sample\cite{Ferrari2007}.

In the past 6 years, there has been a significant step forward in the understanding of Raman spectroscopy in graphene, fuelled by new experimental and theoretical results on doping\cite{Kalbac2010,Chen2011,cinzapl,das1,tandoping,das2,pisana,Yan07,Yan08}, edges\cite{Basko09,You2008,Gupta2009,Cedge,Cong10,Ryu}, strain and stress\cite{heinzinner,maulinner,soninner,Mohi,Ni2008,halsall,Hone}, disorder\cite{Ferrari2007,Venezuela,lucchese,cancadoacf,Ryu}, oxidation\cite{gokus}; hydrogenation\cite{elias}, chemical functionalisation\cite{CF}, electrical mobility\cite{ni,fuhrer}, thermal conductivity\cite{balandin,Bonini2007}, electron-phonon\cite{baskoal,NJP,Basko08,lazzeriM,Venezuela,Bonini2007,Lazzeri08,gruneis2,maul2} and electron-electron\cite{baskoal,Basko08,Lazzeri08,gruneis2}  interactions; magnetic field\cite{Faugeras2009,Kossacki2011,Faugeras2011,smi11,smi12,Faugeras2010,Faugeras2012,pellegrini,falko,Ando07}; interlayer coupling\cite{tan42,Lui1,Lui2,Sato,Maul}. As a result, the understanding of the basic Raman processes has changed. Raman scattering on phonons is to a large extent determined by electrons: how they move, interfere, and scatter. Thus, any variation of electronic properties due to defects or edges, doping, magnetic fields, affects positions, widths, and intensities of the Raman peaks, enabling one to probe electrons via phonons. Quantum interference effects\cite{NJP,Chen2011,janina2} play a key role, and they can also be probed by this technique.

Here we review these new developments, and incorporate them into a general framework for Raman spectroscopy in graphene based on a unified and self-consistent terminology. We introduce the basic physics of Raman scattering in graphene, and discuss the effects of edges, number of layers, defects and disorder, and perturbations. We outline the history of the field, interference\cite{yoonIERS,singa} and surface-enhanced\cite{sers} Raman scattering in the Appendix (S4, S2 and S3, respectively), along with the effects of polarization (S5), electric fields and doping (S6), magnetic field (S7), uniaxial and biaxial strain (S8), temperature (S9), isotopes (S10) and other examples (S11).

The key difference between our framework and those published previously \cite{Malard2009,Jorio2010,Saito2011} is that we start from the general picture of the Raman process, and show how the numerous observed effects naturally arise from it. This approach creates a unified view of Raman scattering, thereby enabling one to better understand the observed effects and, hopefully, anticipate new ones.
\section{The Raman spectrum of graphene}\label{sec:spectrum}
\begin{figure*}
\centerline{\includegraphics [width=180mm]{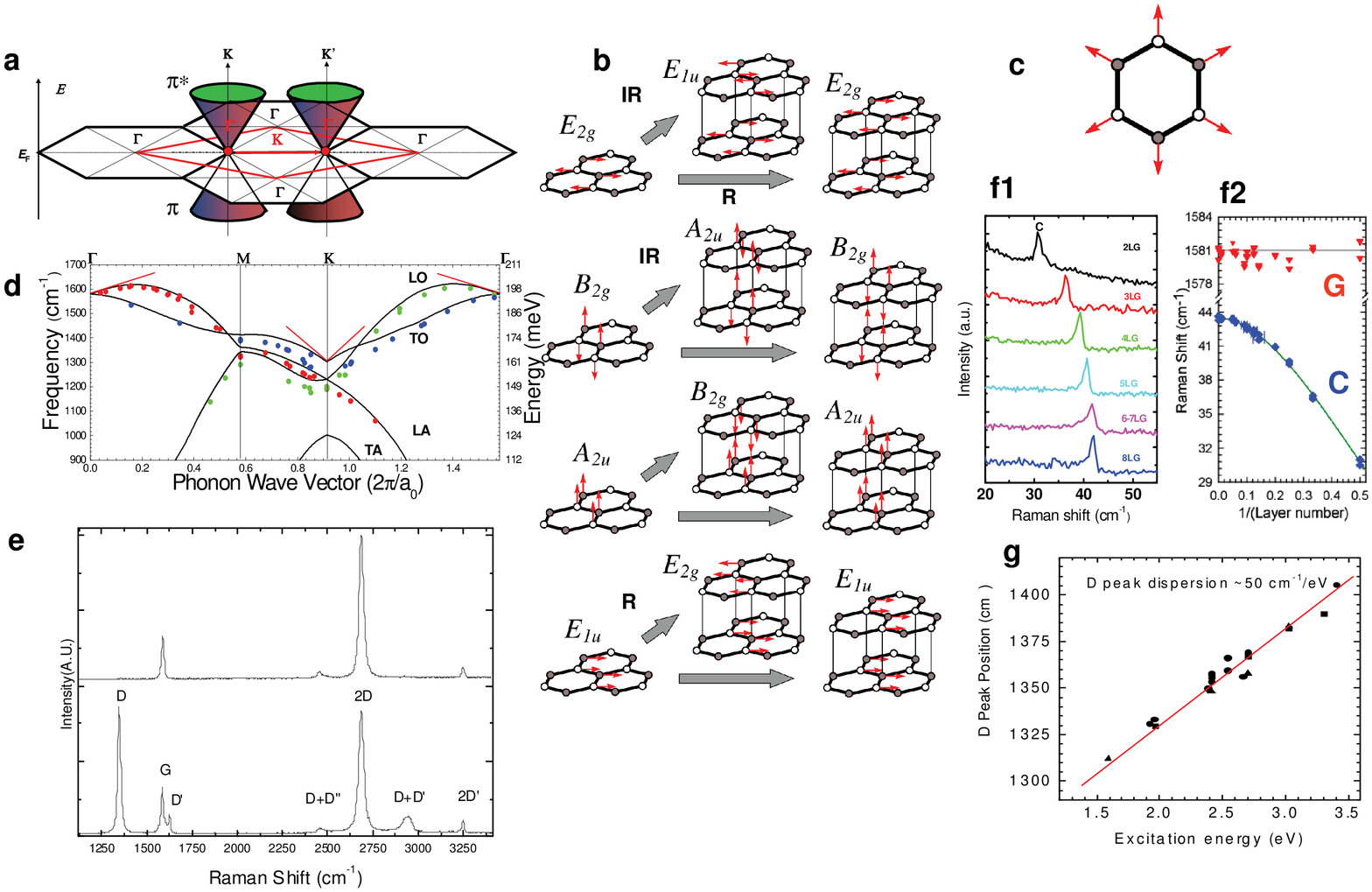}}
\caption{\textbf{Electrons, phonons, and Raman spectrum of graphene.}
\textbf{a,} Electronic BZs of graphene (black hexagons), the first phonon BZ (red rhombus), and schematic electronic dispersion (Dirac cones). The phonon wave vectors connecting electronic states in different valleys are labeled by red Greek letters. \textbf{b}, $\bf{\Gamma}$-point phonon displacement pattern for graphene and graphite. Empty and shaded circles represent inequivalent carbon atoms. Red arrows show atom displacements. Grey arrows show how each phonon mode in graphene gives rise to two phonon modes of graphite. Their labeling shows Raman-active (R), infrared-active (IR) and inactive (unlabeled) modes. \textbf{c}, Atom displacements (red arrows) for the $A_{1g}$ mode at $\mathbf{K}$. \textbf{d}, The black curves represent the dispersion of in-plane phonon modes in graphene in the energy and frequency range relevant for Raman scattering. The red lines represent Kohn Anomalies\cite{piscanec}. The symbols are data taken from Refs.\onlinecite{janina,gruneis2}. \textbf{e}, Raman spectra of pristine (top) and defected (bottom) graphene. The main peaks are labeled. \textbf{f1}, C peak as a function of number of layers.\textbf{f2} Fitted C and G peak positions as a function of inverse number of layers. The line passing through the C peak data is taken from Eq.1. Flakes with N layers are indicated by NLG. Thus, e.g., 2LG is BLG, and 8LG is a 8-layer graphene.\textbf{g}, D peak dispersion with excitation energy, data from Ref.\onlinecite{pocsik}.}
\label{fig1}
\end{figure*}
To understand the state of the art of Raman spectroscopy in graphene it is important to know the historical development of the main ideas, nomenclature and peaks assignment starting from graphite. We thus give a detailed overview in Appendix S4, where we also introduce some background concepts, such as Kohn anomalies\cite{piscanec}. Here we summarize the nomenclature and current understanding of the main peaks.

Throughout this paper we will use the following notation: $I$ for a peak height, $A$ for its area, Pos for its position, FWHM for the full width at half maximum, Disp for its dispersion (i.e. shift in peak position with changing excitation energy). So, e.g., $I(\mathrm{G})$ is the G~peak height, $A(\mathrm{G})$ its area, FWHM(G) the full width, Pos(G) its position, Disp(G) its dispersion.

The phonon dispersion of single-layer graphene (SLG) comprises three acoustic (A) and three optical (O) branches. The modes with out-of-plane (Z) motion are considerably softer than the in-plane longitudinal (L) and transverse (T) ones. Fig.1a plots the Electronic Brillouin Zone (BZ) of graphene, the first phonon BZ, and the schematic electronic dispersion (Dirac cones). Graphene has 2 atoms per unit cell, thus six normal modes at the BZ center $\bf{\Gamma}$ (Refs.\onlinecite{nemanich0,reich1}): $A_{2u}+B_{2g}+E_{1u}+E_{2g}$ (Fig.~1b). There is one degenerate in-plane optical mode, $E_{2g}$, and one out-of-plane optical mode $B_{2g}$ (Ref.\onlinecite{nemanich0}). The $E_{2g}$ phonons are Raman active, while $B_{2g}$ is neither Raman nor Infra Red (IR) active\cite{nemanich0}. Graphite has 4 atoms per unit cell. Only half the carbons have fourth neighbors that lie directly above or below in adjacent layers. Therefore, the two atoms of the unit cell in each layer are now inequivalent. This doubles the number of optical modes, and is responsible for the IR activity of graphite\cite{nemanich0}. All the optical modes become Davydov-doublets: The $E_{2g}$ generates a IR active $E_{1u}$ and a Raman active $E_{2g}$, the $B_{2g}$ goes into an IR-active $A_{2u}$, and an inactive $B_{2g}$. The antisymmetric combinations of the acoustic modes are the optically inactive $B_{2g}$ and the Raman active $E_{2g}$. The symmetric combinations of the acoustic modes remain $A_{2u}$ and $E_{1u}$. Thus, for graphite\cite{mani,reich1,nemanich0} $\Gamma=2(A_{2u}+B_{2g}+E_{1u}+E_{2g})$(Fig.1b). There are now two Raman active $E_{2g}$ modes, each doubly degenerate.

The Raman spectrum of SLG consists of distinct bands\cite{ACFRaman} (Fig.1e). Fig.1d plots the optical phonon dispersions of SLG, relevant for the interpretation of the Raman spectra\cite{piscanec,janina,gruneis2}. The G peak corresponds to the high frequency $E_{2g}$ phonon at $\bf{\Gamma}$. The D peak is due to the breathing modes of six-atom rings (Fig.1c) and requires a defect for its activation\cite{tk,Ferrari00,thomsen}. It comes from TO phonons around the BZ edge \textbf{K} (Refs.\onlinecite{tk,Ferrari00}), is active by double resonance\cite{thomsen,baranov}, and is strongly dispersive with excitation energy\cite{pocsik} (Fig.1g), due to a Kohn anomaly at \textbf{K} (Ref.\onlinecite{piscanec}). Double resonance can also happen as intra-valley process, i.e. connecting two points belonging to the same cone around $\textbf{K}$ (or $\textbf{K}'$). This gives the so-called D$'$ peak. The 2D peak is the D peak overtone. The 2D$'$ peak is the D$'$ overtone. Since 2D and 2D$'$ originate from a process where momentum conservation is satisfied by two phonons with opposite wavevectors, no defects are required for their activation, and are thus always present\cite{ACFRaman,BPF09}.

The band at $\sim{2}450\:\mbox{cm}^{-1}$ in Fig.~1e was first reported in graphite by Nemanich and Solin\cite{nema3}. Its interpretation was subject to debate, as discussed in Appendix S4. It is assigned to a combination of D~phonon and a phonon belonging to the LA branch, seen at $\sim{1}100\:\mbox{cm}^{-1}$ for visible excitation in defected samples, and called D$''$ peak\cite{kava,tan1,tan2,tan3,maul2}. It is indicated as $\mbox{D}+\mbox{D}''$ in Fig.~1e.

The Raman spectrum of graphite and multi-layer graphene consists of two fundamentally different sets of peaks. Those, such as D, G, 2D and so on, present also in SLG, and due to in-plane vibrations\cite{ACFRaman,acftrans,tk,Ferrari2007}, and others, such as the shear (C) modes\cite{tan42} and the layer breathing modes (LBMs)\cite{Lui1,Lui2,Sato}, due to relative motions of the planes themselves, either perpendicular or parallel to their normal. The low-frequency $E_{2g}$ mode in graphite was first measured by Nemanich \textit{et al.} in 1975\cite{nema1} at $\sim42\:\mbox{cm}^{-1}$. We called this mode C, since it is sensitive to the interlayer coupling\cite{tan42}  (Fig.~1f). The absence of the C peak would in principle be the most direct evidence of SLG. However, it is not warranted to use the absence of a peak as a characterization tool, since one can never be sure why something is absent. On the other hand, this mode scales with the number of layers ($N$), going to $\sim31\:\mbox{cm}^{-1}$ for bilayer graphene\cite{tan42} (BLG) (Fig.~1f). The C peak frequency is below the notch and edge filter cut of many spectrometers, particularly those used for production-line monitoring. This problem was overcome recently by combining a BragGrate filter with a single monochromator\cite{tan42} and, for the first time, Pos(C) was measured for an arbitrary $N$ (Ref.~\onlinecite{tan42}). This method allows detection of similar modes in any other layered material\cite{tan42,tanmos2}. For Bernal stacked samples, Pos(C) varies with $N$ as\cite{tan42}:
\begin{equation}
\mbox{Pos}(\mbox{C})_N=\sqrt{\frac{2\alpha}{\mu}}\sqrt{1+\cos\Big(\frac{\pi}{N}\Big)}
\label{eq:freq2}
\end{equation}
where $\alpha=12.8\times 10^{18}\:\mbox{N/m}^3$ is the interlayer coupling and $\mu=7.6\times 10 ^{-27}\:\mbox{kg/\AA}^2$ is the graphene mass per unit area\cite{tan42}. LBMs can also be observed in the Raman spectra of FLGs, via their resonant overtones in the range $80-300\:\mbox{cm}^{-1}$ (Refs.~\onlinecite{Lui1,Lui2,Sato,Maul}). Note that, albeit being an in-plane mode, the 2D peak is sensitive to $N$ since the resonant Raman mechanism that gives rise to it is closely linked to the details of the electronic band structure\cite{ACFRaman,Ferrari2007}, the latter changing with $N$, and the layers relative orientation\cite{latil}. On the other hand, the C and LB modes are a direct probe of $N$ (Refs.~\onlinecite{tan42,Lui1,Lui2,Sato,Maul}), as the vibrations themselves are out of plane, thus directly sensitive to $N$. Because the fundamental LBM in bulk graphite is a silent $B_{1g}$ mode at $\sim128\:\mbox{cm}^{-1}$, the observation of the first-order LBM is a challenge.

Raman spectroscopy can also probe scattering of photons by electronic excitations. In pristine graphene, these have a continuous structureless spectrum\cite{wallace}, not leading to any sharp feature. However, it was realized that in a strong magnetic field, $B$, when the electronic spectrum consists of discrete Landau levels, the electronic inter-Landau-level excitations give rise to sharp $B$-dependent peaks in the Raman spectrum\cite{Kashuba2009,Kashuba2012,Faugeras2011,smi11,smi12}.
\section{Raman processes in graphene}\label{sec:processes}
The understanding of Raman spectroscopy in graphite and related materials has challenged researchers for decades. The reason is the richness of phenomena combined with the wealth of experimental information that must be consistently arranged to solve the jigsaw. An introduction to Raman scattering is presented in Appendix S1, whereas surface-enhanced and interference-enhanced Raman spectroscopy in graphene are discussed in Appendixes S2,S3.
\begin{figure*}
\centerline{\includegraphics [width=180mm]{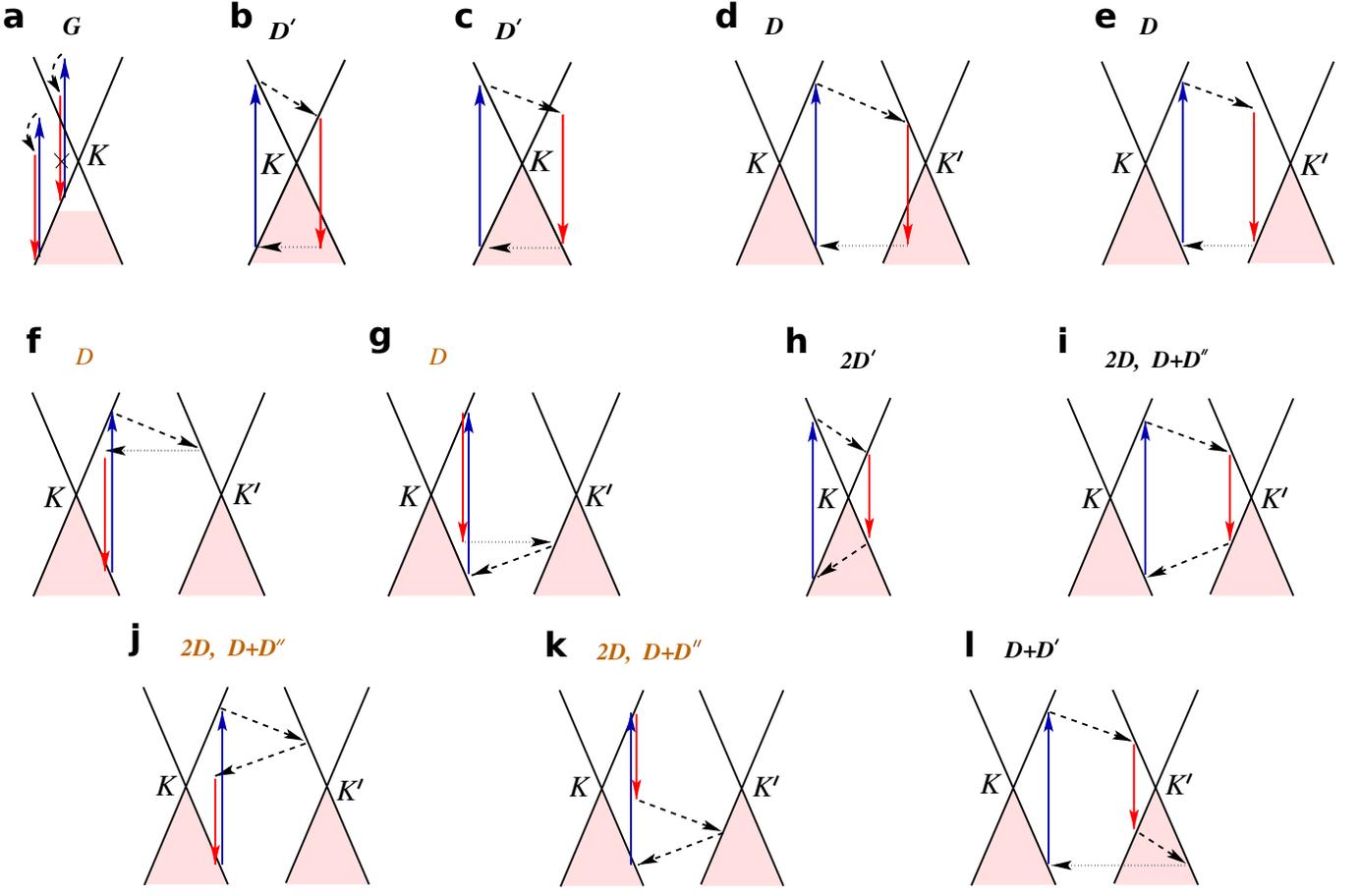}}
\caption{\label{fig:res}
\textbf{Raman processes}. Electron dispersion (solid black lines), occupied states (shaded areas), interband transitions neglecting the photon momentum (vertical solid arrows), photon absorption (blue lines) and  emission (red lines), intraband transitions accompanied by phonon emission (dashed arrows), electron scattering on a defect (horizontal dotted arrows). \textbf{a}~One-phonon processes responsible for the G peak, which interfere destructively. Some processes can be eliminated by doping, such as the one which is crossed out. \textbf{b--g} In the presence of defects, the phonon wave vector need not be zero, producing the D$'$ peak for intravalley scattering (\textbf{b,c}), and D peak for intervalley scattering (\textbf{d--g}). Besides the $eh$ or $he$ processes, where the electron and the hole participate in one act of scattering each (\textbf{b--e}), there are contributions $ee$, $hh$, where only the electron (\textbf{f}) or the hole~(\textbf{g}) are scattered. For two-phonon scattering, momentum can be conserved by emitting two phonons with opposite wavevectors, producing the 2D$'$ peak for intravalley scattering (\textbf{h}) and the 2D, D+D$''$ peaks for intervalley scattering (\textbf{i--k}). \textbf{j,k} show the $ee$ and $hh$ processes. \textbf{l} With defects, one intravalley and one intervalley phonon can be emitted, producing the D+D$'$ peak. The processes \textbf{f,g,j,k} give a small contribution, as indicated by orange peak labels.}
\end{figure*}

In graphene, graphite and nanotubes, Raman processes involving up to six phonons can be easily measured\cite{heinz,tan1,tan2,tan3,Rao2011}. However, the vast majority of literature reports spectra up to $\sim3300\:\mbox{cm}^{-1}$. This restricts our attention to one-and two-phonon peaks. We can also distinguish between spectra measured on pristine samples (i.e., ideally defect-free, undoped, unstrained and so on) and those measured on samples subject to external perturbations (such as electric and magnetic fields, strains and so on) or those with defects. We will cover external perturbations below, as well as in Appendixes S6-S9. Defect-activated peaks will be discussed now together with those not requiring defects for their activation.

In general, Raman scattering can be described by perturbation theory\cite{YuCardona}. An $n$-phonon process involves $n+1$ intermediate states, and is described by an $(n+2)$-order matrix element, as given by Equation (S3) in Appendix. Figure 2 plots the possible elementary steps of the Raman processes contributing to each peak of graphene. According to the number of factors
in the denominator of Eq.~(S3) that vanish, the processes can be classified as double resonant (Fig.~2b-g,j,k) or triple resonant (Fig.~2h,i,l). Higher orders are also possible in multiphonon
processes. Note that this classification is useful, but approximate (valid when the electron-hole asymmetry and the difference in energies of the two phonons are neglected). The process in Fig.2a is not resonant.

One-phonon modes in defect-free samples can be Raman active only if their symmetry is correct and their wavevector is zero (i.e., they obey the fundamental Raman selection rule, see Appendix S1). In graphene, only the C and G peaks satisfy these requirements. The energies of the intermediate states are given by the difference in energies of electrons in the empty $\pi^*$ and filled $\pi$ bands, $\ep^{\pi^*}_\vec{k}-\ep^{\pi}_\vec{k}$ (with $\vec{k}$ the electronic wave vector), with or without the phonon energy, $\hbar\Omega_{\vec{q=0}}$ at $\vec{q}=0$. The decay rate of the intermediate states is given by the sum of the scattering rates of the electron in the $\pi^*$~band, $2\gamma^{\pi^*}_\vec{k}/\hbar$, and of the hole in the $\pi$~band, $2\gamma^{\pi}_\vec{k}/\hbar$. The contribution from the phonon decay is typically smaller.

Counterintuitively, the electronic wave vectors $\vec{k}$ mostly contributing to the matrix element for the G peak are not only those for which the excitation energies $\ep^{\pi^*}_\vec{k}-\ep^{\pi}_\vec{k}$ lie within an interval $\sim\gamma$ from $\hbar\omega_L$ or $\hbar\omega_L-\hbar\Omega_{\vec{q}=0}$, where $\hbar\omega_L$ is the incident laser photon energy. Instead, they are such that $|\ep^{\pi^*}_\vec{k}-\ep^{\pi}_\vec{k}-\hbar\omega_L|$ can be of the order of $\hbar\omega_L$ itself, and there are strong cancellations in the sum over~$\vec{k}$ (Ref.\onlinecite{NJP}). These correspond to destructive quantum interference. In fact, this interference can be controlled externally. Indeed, occupations of electronic states can be changed by doping and, since transitions from an empty state or to a filled state are impossible due to Pauli blocking, doping can effectively exclude some regions of~$\vec{k}$ from contributing to the matrix element (Fig.~2a). Due to suppression of destructive interference, this leads to an increase of the G peak intensity at high doping levels, as was predicted in Ref.~\onlinecite{NJP} and observed in Refs.~\onlinecite{Kalbac2010,Chen2011}.

For two-phonon processes, the fundamental selection rule can be obeyed by any pair of phonons with opposite wave vectors, $\vec{q},-\vec{q}$. The matrix element has four contributions corresponding to processes when (i)~both phonons are emitted by the electron (${ee}$), (ii)~both phonons are emitted by the hole ($hh$), and (iii)~one phonon is emitted by the electron, and the other by the hole ($eh$ and $he$).
In principle, one could expect the two-phonon Raman spectrum to be a broad band, as determined by the sum of phonon frequencies $\Omega^\alpha_\vec{q}+\Omega^\beta_{-\vec{q}}$ from branches $\alpha,\beta$ for all~$\vec{q}$, with possible features from van Hove singularities in the joint phonon density of states. However, resonance conditions favor phonon states with $\vec{q}$ coupling electronic states $\vec{k}$, $\vec{k}-\vec{q}$, either in the same valley (i.e. with $\vec{q}$ near $\bf{\Gamma}$), or in different valleys ($\vec{q}$ near \textbf{K}). But, it turns out that even amongst these $\vec{q}$, very few are selected by subtle effects of resonance and quantum interference\cite{Basko08,Venezuela}. These effects are, of course, captured by the direct integration over the electron momentum, where they appear as cancelations in the sum over~$\vec{k}$, but they are most easily understood by considering the Raman process in real space.

\begin{figure*}
\centerline{\includegraphics [width=180mm]{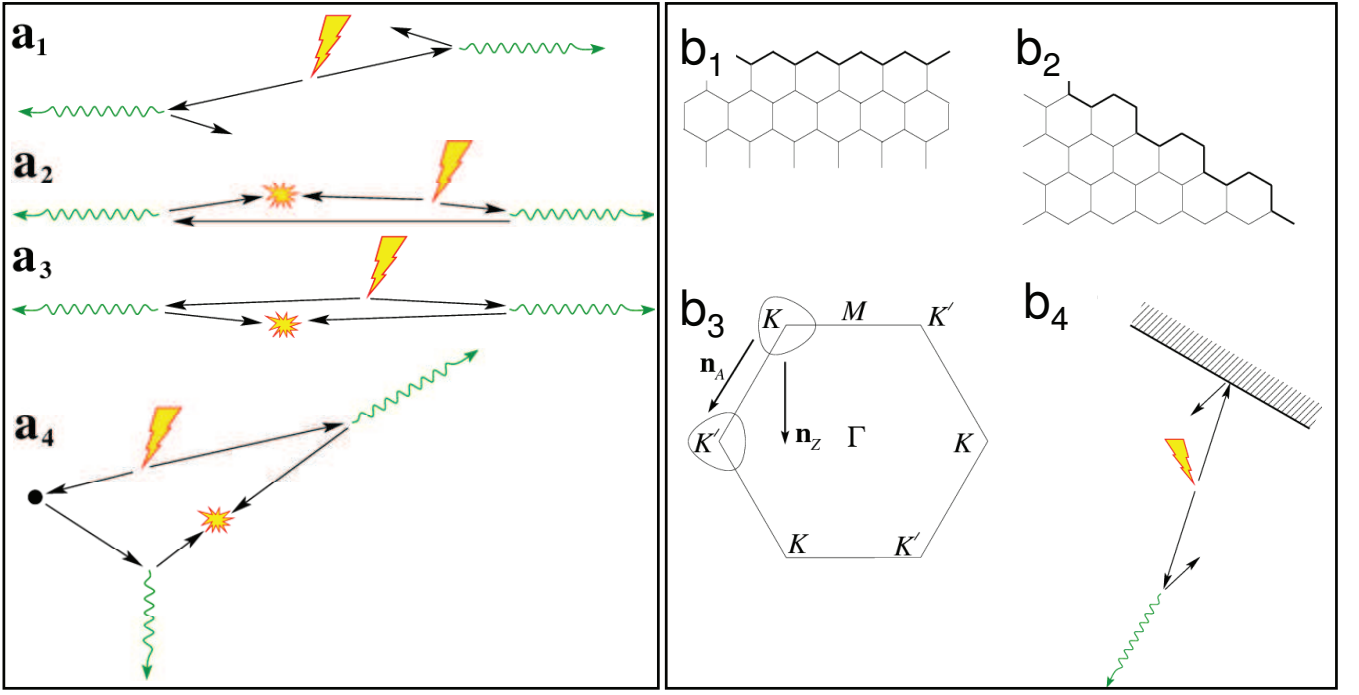}}
\caption{\label{fig:traject}
\textbf{Real-space Raman processes}. The excitation photon promotes an electron with momentum $\vec{p}=\hbar\vec{k}$ from $\pi$ to $\pi^*$, thereby creating a hole in the $\pi$ band with momentum $-\vec{p}$, and energy $-\ep^\pi_\vec{k}$, as shown by the lightning. This process is assumed to take place in a given point in space. $e$ and $h$ then move along classical trajectories, in directions determined by their group velocities, $\vec{v}^e_\vec{k}=\partial\ep^{\pi^*}_\vec{k}/\partial(\hbar\vec{k})$, $\vec{v}^h_\vec{k}=\partial\ep^{\pi}_\vec{k}/\partial(\hbar\vec{k})$, as shown by straight solid arrows. At some point, they emit phonons (shown by wavy lines) or scatter on defects (shown by dots) or edges (hatched zones). To recombine radiatively and produce the scattered photon (flash), $e$ and $h$ must meet with opposite momenta $\vec{k}',-\vec{k}'$ at the same point in space, after having traveled for the same amount of time. \textbf{(a)}, Trajectories for two-phonon processes. $\bf{a_1}$ Trajectory for which radiative recombination is impossible, even though momentum is conserved, because $e$ and $h$ cannot meet at the same point to recombine. $\bf{a_2}$, trajectory corresponding to an $ee$ process, incompatible with the requirement that $e$ and $h$ travel for the same amount of time. $\bf{a_3}$, Trajectory corresponding to 2D, 2D$'$. Upon phonon emission, $e$ and $h$ must be back-scattered. $\bf{a_4}$, Trajectory corresponding to $\mathrm{D}+\mathrm{D}'$. \textbf{(b)},~Scattering at edges. $\bf{b_1}$, Zigzag edge. $\bf{b_2}$, Armchair edge. $\bf{b_3}$, Equi-energy contours for electronic states involved in the D peak. $\vec{n}_Z$ and $\vec{n}_A$ indicate directions normal to zigzag and armchair edges, respectively. $\bf{b_4}$, A trajectory not contributing to the D peak, as $e$ and $h$ cannot meet after scattering.}
\end{figure*}

Raman scattering in the real space was first considered in 1974\cite{Zeyher74}, and the spatial separation between electron and hole in cascade multiphonon Raman scattering was analyzed in 1983\cite{Goltsev83}. However, in the context of graphene this approach was proposed only recently\cite{Basko08,Cedge,Basko09}. The real-space picture is especially useful when translational invariance is lacking due to defects or edges. It arises because of separation of two energy scales: the electronic energy $\ep\approx\hbar\omega_L/2$ ($\sim1\:\mbox{eV}$ for visible Raman), and the energy uncertainty $\delta\ep\ll\ep$. For the triple-resonant processes in Fig.~\ref{fig:res}h,i,l, $\delta\ep$ is of the order of the broadening $\gamma$ (a few tens meV; Refs.\onlinecite{BPF09,Faugeras2010,Venezuela}). For the double-resonant processes in Fig.~\ref{fig:res}b--g,j,k, $\delta\ep$ is of the order of the phonon energy $\hbar\Omega$, 0.17 or 0.20~eV for phonons near $\bf{\Gamma}$ or $\vec{K},\vec{K}'$ (Ref.~\onlinecite{piscanec}). From the uncertainty principle, $\delta\ep$ determines the typical lifetime of the intermediate state, $\sim\hbar/\delta\ep$, be it real or virtual. This gives the process duration, while $\ell=\hbar{v}_F/\delta\ep$ its spatial extent ($v_F\approx{10}^8\:\mbox{cm/s}\approx{7}\:\mbox{eV}\cdot\mbox{\AA}\,\hbar^{-1}$ is the Fermi velocity). For triple-resonant processes, $\gamma=20\:\mbox{meV}$, we have $\ell\sim\hbar{v}_F/\gamma\approx{35}\:\mbox{nm}$; for double-resonant processes, $\ell\sim{v}_F/\Omega\approx{3}.5\:\mbox{nm}$ (for $\hbar\Omega=0.2\:\mbox{eV}$). Since $\ell$ is much bigger than the electron wavelength, $\lep=\hbar{v}_F/\ep\approx{0}.7\:\mbox{nm}$ for $\hbar\omega_L=2\:\mbox{eV}$, the electron and hole motion can be viewed in quasi-classical manner, as shown in Fig.\ref{fig:traject} for two-phonon processes. This is analogous to the geometrical optics approximation for electromagnetic waves, with electronic trajectories corresponding to light rays. The quasi-classical picture arises when calculating the Raman matrix elements in the coordinate representation\cite{Basko09}. It does not require real $e,h$ populations, as it is a property of wave functions.

From the real-space picture it is seen that two-phonon processes are mostly contributed by those $\vec{q}$ which correspond to $e$ and $h$ scattered backwards, otherwise $e$ and $h$ will not meet in the same point. The backscattering condition, corresponding to reversal of the direction of the group velocity, agrees with the nesting condition of Ref.\onlinecite{Venezuela}, and has been mentioned as early as in 1974\cite{martin}. In particular, this eliminates the contribution from two phonons with $\vec{q}=0$ which would correspond, e.g., to the G peak overtone, explaining why the 2G peak is not seen in the Raman spectrum (unfortunately many works still mistakenly name the 2D$'$ peak as 2G). Also, the $ee$ and $hh$ processes, where one of the carriers has to travel longer than the other, are in conflict with the requirement that they travel for the same amount of time (this would not be the case if $e$ and $h$ had strongly different velocities). Strictly speaking, such processes, prohibited in the classical picture (corresponding to $\delta\ep/\ep\to{0}$), are allowed in the full quantum picture, but they are weaker than the dominant contribution. Note that the G peak is also classically forbidden ($e$ and $h$ cannot meet at the same point). Thus, one may wonder why it produces a noticeable feature in the spectrum, while classically forbidden processes in two-phonon scattering are not seen. One reason is that two-phonon processes contain a higher power of the small electron-phonon coupling (EPC); another is that the G peak is narrow (a few cm$^{-1}$), while the two-phonon bands spread over hundreds of cm$^{-1}$.

In samples with defects the overall momentum conservation can be satisfied by adding an electron-defect scattering event to the process. We can thus have (i)~one-phonon defect-assisted processes, producing the D, D$'$, D$''$ and other smaller peaks, (ii)~two-phonon defect-assisted processes, such as those leading to the $\mathrm{D}+\mathrm{D}'$ peak. For one-phonon defect-assisted processes the matrix element has the same form as two-phonon defect-free processes, it is sufficient to replace the electron-phonon Hamiltonian  by the electron-defect Hamiltonian, and set the frequency of the second phonon to zero. For the two-phonon defect-assisted processes the situation is quite different. The explicit formula for the matrix element contains 48 terms, each having five matrix elements in the numerator and a product of four factors in the denominator. Fig.\ref{fig:traject}a$_4$ shows that for the $\mathrm{D}+\mathrm{D}'$ peak there is no backscattering restriction, so the phonon momenta (counted from \textbf{K} and $\bf{\Gamma}$) can be between 0 and $\omega_L/(2v_F)$. Thus the $\mathrm{D}+\mathrm{D}'$ peak is much broader than 2D or 2D$'$, since in the latter the magnitude of the phonon momentum is fixed by the backscattering condition, and $\mbox{Pos}(\mathrm{D}+\mathrm{D}')\neq\mbox{Pos}(\mathrm{D})+\mbox{Pos}(\mathrm{D}')$. Unfortunately, the $\mathrm{D}+\mathrm{D}'$ peak is often labeled as D+G in literature, due to lack of understanding of this activation mechanism.

These simple considerations can explain the peaks' dispersions. E.g., in the 2D process (Fig.~\ref{fig:res}i), the photon creates $e$ and $h$ with momenta $\vec{p}$ and $-\vec{p}$, counted from the Dirac points. These then emit phonons with momenta $\hbar\vec{q}=\vec{p}-\vec{p}'$ and $-\hbar\vec{q}$ (counted from the $\vec{K}$ point). In the Dirac approximation for the electronic dispersion, $\ep^{\pi^*}_{\vec{K}+\vec{p}/\hbar}=-\ep^{\pi}_{\vec{K}+\vec{p}/\hbar}=v_F|\vec{p}|$, and assuming isotropic phonon dispersions around $\vec{K}$, the magnitudes of $\vec{p}$ and $\vec{p}'$ before and after phonon emission are fixed by the resonance condition: $2v_Fp=\hbar\omega_L$, $2v_Fp'=\hbar\omega_L-2\hbar\Omega^{TO}_q$. The backscattering condition (Fig.~\ref{fig:traject}a) fixes the relative orientation of $\vec{p}$,$\vec{p}'$ (i.e., opposite directions), thus the magnitude of $\vec{q}$ is:
\begin{equation}\label{q2D=}
{q}=\frac{p+p'}\hbar=\frac{\omega_L}{2v_F}+\frac{\omega_L-2\Omega^{TO}_q}{2v_F}
\end{equation}
Neglecting $\Omega^{TO}_q$, we estimate Disp(2D) as:
\begin{equation}
\mbox{Disp}(2\mbox{D})=\frac{d\mathrm{Pos}(2\mbox{D})}{d\omega_L}=2\,\frac{d\Omega^{TO}_q}{dq}\,\frac{dq}{d\omega_L}\approx
{2}\,\frac{v^{TO}_q}{v_F},
\end{equation}
where $v^{TO}_q\equiv{d}\Omega^{TO}_q/dq$ is the TO phonon group velocity. Since $d\mathrm{Pos}(2\mathrm{D})/d\omega_L\approx{1}00\:\mathrm{cm}^{-1}/\mathrm{eV}$ (Ref.\onlinecite{vidano2}), we obtain $v^{TO}_q\approx{0}.006\,v_F$. A similar argument holds for the $\mathrm{D}+\mathrm{D}''$ peak, whose experimentally observed dispersion$\approx-{20}\:\mathrm{cm}^{-1}/\mathrm{eV}$ is determined as $v^{TO}_q/v_F+v^{LA}_q/v_F$ (the LA phonon has a negative slope as the wave vector moves away from the \textbf{K} point)\cite{mafra,maul2}.

An understanding of the main contributions to the widths of the Raman peaks in graphene has been recently achieved. Besides the anharmonic and EPC terms, which dominate the width of the G peak\cite{pisana,Yan07,Bonini2007}, other factors can contribute to the widths of higher-order peaks. One is the $e$, $h$ dispersion anisotropy, since phonons emitted by $e$ with different $\vec{p}$ directions have different frequencies\cite{Venezuela}. Another is the uncertainty in momentum due to the electronic scattering rate\cite{Basko08,Faugeras2010}. When all these mechanisms are taken into account (note that they are not simply additive), the experimental width of the peaks, such as D and 2D, is reproduced\cite{Venezuela}.

The intensities of the two-phonon peaks are suppressed by electronic scattering\cite{Basko08,Venezuela}. Anisotropy of the latter\cite{Basko2013,Golub2011}, together
with the EPC anisotropy\cite{Venezuela}, makes the electronic states close to the $\mathbf{K}-\mathbf{M}$ line (the so-called inner resonance) dominate the two-phonon Raman process for visible excitation, which explains recent experimental observations\cite{heinzinner,pimentainner,soninner}. Different contributions come to play when moving towards UV excitation.

For the D and D$'$ peaks, the electron and hole momenta can have different magnitudes, depending on which of the intermediate states is virtual\cite{cancado1}, as shown in Fig.~\ref{fig:res}d,e. Arguments analogous to those presented above give two possibilities for the emitted phonon momentum: $q=\omega_L/v_F$ and $q=(\omega_L-\Omega^\alpha_q)/v_F$. This means that D could in principle have two components\cite{cancado1} separated by $(v_q^{TO}/v_F)\Omega_q^{TO}\approx{8}\:\mathrm{cm}^{-1}$ (this should not be confused with the two D peak components seen in graphite\cite{nema3}, which reflect the three-dimensional electron dispersion, as discussed in Section~\ref{sec:layers}\cite{ACFRaman}, nor with the broadening and splitting expected for UV excitation). Such splitting, if present, should be seen in the Raman spectrum. However even in SLG with very few defects, $\mbox{FWHM}(\mathrm{D})\approx{20}\:\mathrm{cm}^{-1}$, thus larger than this splitting. This also means that $\mbox{Pos}(\mathrm{D})$ can be slightly higher than $(1/2)\mbox{Pos}(2\mathrm{D})$, as the latter is uniquely determined by $q_{2\mathrm{D}}=(\omega_L-\Omega_q^{TO})/v_F$. For D$'$, $(v_q^{LO}/v_F)\Omega_q^{LO}\sim{2}\:\mbox{cm}^{-1}$, thus the difference between $\mbox{Pos}(2\mathrm{D}')$ and $2\,\mbox{Pos}(\mathrm{D}')$ is barely detectable.

The intensities of different peaks were reported to depend differently on the excitation frequency $\omega_L$ (Ref.~\onlinecite{Cancado07}): while $A(\mathrm{D}),A(\mathrm{D}'),A(2\mathrm{D})$ show no $\omega_L$-dependence, $A(\mathrm{G})$ was found to be $\propto\omega_L^4$. The independence of $A(2\mathrm{D})$ on $\omega_L$ agrees with the theoretical prediction\cite{Basko08} if one assumes that the electronic scattering rate is proportional to the energy. For the G peak, Ref.~\onlinecite{NJP} predicted $A(G)\propto\omega_L^2$ at low~$\omega_L$, and strong enhancement of $A(G)$ as $\omega_L$ approaches the van Hove singularity corresponding to the electronic excitation at the \textbf{M} point, reported to be as low as 4.6~eV (Refs.~\onlinecite{Yang2009,Kravets2010}). It cannot be excluded that the $\omega_L^4$-dependence, established empirically from measurements in the range $\hbar\omega_L=1.6-2.7\:\mbox{eV}$, corresponds to the crossover between $\omega_L^2$ and the enhancement at the van Hove singularity. For the D peak, Ref.~\onlinecite{Basko09} gave $A(\mathrm{D})\propto\omega_L$. The reason for the discrepancy with experiments remains unclear. Note, however, that a fully quantitative theory for $A(\mathrm{D})$ and its excitation energy dependence is not trivial since, in general, $A(\mathrm{D})$ depends not only on the concentration of defects, but on their type as well (e.g., only defects able to scatter electrons between the two valleys can contribute). Different defects can also produce different frequency and polarization dependence of $A(\mathrm{D})$. Thus a wealth of information still remains to be uncovered from the D peak.

So far, we only considered Stokes processes. For anti-Stokes, the resonance conditions are modified. For the 2D peak, the momentum of the absorbed phonons is determined by an equation similar to Eq.~(\ref{q2D=}), but with a ``+'' sign in front of $2\Omega^{TO}_q$. Thus, the phonon momenta in the anti-Stokes process are greater than those in the Stokes process by $2\Omega_q^{TO}/v_F$. This explains the Stokes/anti-Stokes process difference\cite{tan2,tan3,cancado1}:
\begin{equation}
\mbox{Pos}(2D)_\mathrm{AS}-\mbox{Pos}(2D)_\mathrm{S}\approx
4\Omega_q^{TO}\,\frac{v_q^{TO}}{v_F}\approx{34}\:\mbox{cm}^{-1}.
\end{equation}
As discussed above, the D, D$'$ peaks could, in principle, have two components. In this case, one of them has the same frequency in the Stokes and anti-Stokes scattering, while the other corresponds to a lower frequency in Stokes, and an higher one in anti-Stokes (Ref.~\onlinecite{cancado1}).

It is important to consider the Stokes/anti-Stokes difference when determining the temperature $T$ from the Stokes/anti-Stokes ratio, both in nanotubes and graphene, as the phonons involved in Stokes and anti-Stokes are not the same, unlike what generally happens in other materials, and contrary to what
is required for determining $T$ from this ratio. Accurate calibra tions are thus needed. It should come as no surprise if a mismatch
between expected and measured $T$ arises.
\section{Edges}
Edges are preferred sites to attach functional groups\cite{cervantes}, and their electronic and magnetic properties are different from the bulk\cite{cervantes}. Figs.~3b$_1$,b$_2$ show zigzag and armchair edges schematically. There is evidence that such ideal edges can be produced by chemical\cite{biro} or thermal treatments\cite{girit}. However, most exfoliated samples, even if with macroscopically smooth edges oriented at well-defined angles, are in fact not necessarily microscopically ordered, with disordered edges, often consisting of zigzag and armchair segments\cite{Kobayashi05,girit}.

Edges can be viewed as extended defects, breaking the translational symmetry, even though a perfect zigzag or armchair edge still preserves the translation symmetry along the edge. An immediate consequence of symmetry breaking is the D,D$'$ peak activation. However, a closer look at the Raman process shows that a perfect zigzag edge cannot produce the D~peak\cite{Cancado04,You2008,Cedge}. Due to translational invariance along the edge, a perfect zigzag edge cannot scatter electrons between the $\mathbf{K}$ and $\mathbf{K}'$ valleys, Fig.3b$_3$. This does not apply to D$'$. This peak, not involving intervalley scattering, can be activated both by armchair and zigzag edges, since both $\vec{n}_Z$, $\vec{n}_A$  are compatible with the required intravalley scattering. This is not enough to determine whether one type of edge is more efficient than the other for D$'$-peak activation.

In the real space, a perfect armchair edge requires the electronic momentum to be perpendicular to it, in order to contribute to D, otherwise $e$ and $h$ cannot meet and recombine radiatively (Fig.~3b$_4$). Thus, $\vec{k}$ needs to be along the $\bf{\Gamma}-\vec{K}-\vec{M}$ line, in order to contribute to D~peak. This, in combination with the $[\vec{e}\times\vec{k}]$ dependence for the dipole transition matrix element ($\vec{e}$ being the polarization vector), discussed in Appendix S5, results in a strong polarization dependence: $I(D)\propto\cos^2\theta$, if the excitation is performed with linearly polarized light, oriented at an angle~$\theta$ with respect to the edge, and all polarizations are detected for the scattered light, or $I(D)\propto\cos^4\theta$ if an analyzer, parallel to the polarizer, is used. For D' the real-space picture is analogous to that for D. For regular edges this leads to the same  $I(\mathrm{D}')\propto\cos^2\theta$ or $\cos^4\theta$ dependence, as observed in Refs.~\onlinecite{Cancado04,You2008,Gupta2009,Cedge}.

For a disordered edge, $I(\mathrm{D})$ and $I(\mathrm{D}')$ do not vanish even for $\theta=\pi/2$ (polarization $\vec{e}$ perpendicular to the edge). Indeed, the polarization dependence for a disordered edge, is determined by contributions from armchair and zigzag segments with different orientations. The ratio $I(\mathrm{D})_{min}/I(\mathrm{D})_{max}$ can be used as a measure of edge imperfection\cite{Cedge}.

The G peak, even though present in the bulk, can be modified near an edge. The edge changes the electronic states, which affects (i)~the EPC correction to the phonon frequency, and (ii) electron-phonon and electron-photon matrix elements, responsible for the Raman process itself\cite{Sasaki09}. The correction to the phonon frequency vanishes for displacements perpendicular to the zigzag or armchair edge. The electron-phonon and electron-photon matrix elements vanish when displacements or light polarization are parallel (perpendicular) to the zigzag (armchair) edge. Thus, one of the two modes (depending on edge type) does not show a Kohn anomaly, and does not contribute to the Raman spectrum. This gives (i) position-dependent $\mbox{Pos}(G)$\cite{Sasaki09}; (ii) strong polarization dependence of I(G) near the edge: $I(G)\propto\cos^2\theta$ for a perfect armchair and $I(G)\propto\sin^2\theta$ for a perfect zigzag edge, allowing the edge type to be probed\cite{Cong10} and even controlled\cite{Begliarbekov}.
\section{Number and relative orientation of layers}\label{sec:layers}
There is a significant change in shape and intensity of the 2D peak when moving from SLG to graphite\cite{ACFRaman} (Fig.~4a). The same holds for the D peak, as shown in Fig.~4b, where the D peak at an SLG edge is compared to that at a graphite edge. The 2D peak in graphite has two components $2D_1$ and $2D_2$ (Refs.~\onlinecite{vidano2,nema3}), roughly $1/4$ and $1/2$ of $I(\mathrm{G})$. Fig.4a plots the evolution of the 2D peak as a function of $N$ for 514.5 and 633nm excitations\cite{ACFRaman}. Fig.4c shows that BLG has a different 2D peak compared to both SLG and graphite\cite{ACFRaman}. It has four components, $2D_\mathrm{1B}$, $2D_\mathrm{1A}$, $2D_\mathrm{2A}$, $2D_\mathrm{2B}$, two of which, $2D_\mathrm{1A}$ and $2D_\mathrm{2A}$, have higher intensities. For $N>5$ the spectrum is hardly distinguishable from graphite. We note that Ref.\onlinecite{Faugeras2012} used the evolution of the G peak in a strong magentic field to identify a four-layer sample.

The evolution of the electronic bands with $N$ explains the 2D peak change \cite{ACFRaman}. In BLG, the interaction of the graphene planes causes the $\pi$ and $\pi^*$ bands to divide in four, with a different splitting for $e$ and $h$ (Fig.~4d). The incident light couples only two pairs of the four bands, (Fig.~4d). On the contrary, the two almost degenerate TO phonons can couple all bands. The resulting four processes involve phonons with momenta $\vec{q}_\mathrm{1B}$, $\vec{q}_\mathrm{1A}$, $\vec{q}_\mathrm{2A}$, $\vec{q}_\mathrm{2B}$ (Fig.~4d). $\vec{q}_{\rm 1A}$, $\vec{q}_{\rm 2A}$ link bands of the same type and are associated to processes more intense than $\vec{q}_{\rm 1B}$, $\vec{q}_{\rm 2B}$, since the portion of the phase space where triple resonance is satisfied is larger. These correspond to different frequencies, due to the phonon dispersion around \textbf{K} (Fig.~1d) (Ref.\onlinecite{piscanec}). This gives four peaks in BLG\cite{ACFRaman}. Their excitation-energy dependence is determined by both electron and phonon dispersions. Measuring it probes the interlayer coupling for both\cite{malard,mafra}.
\begin{figure*}
\centerline{\includegraphics [width=180mm]{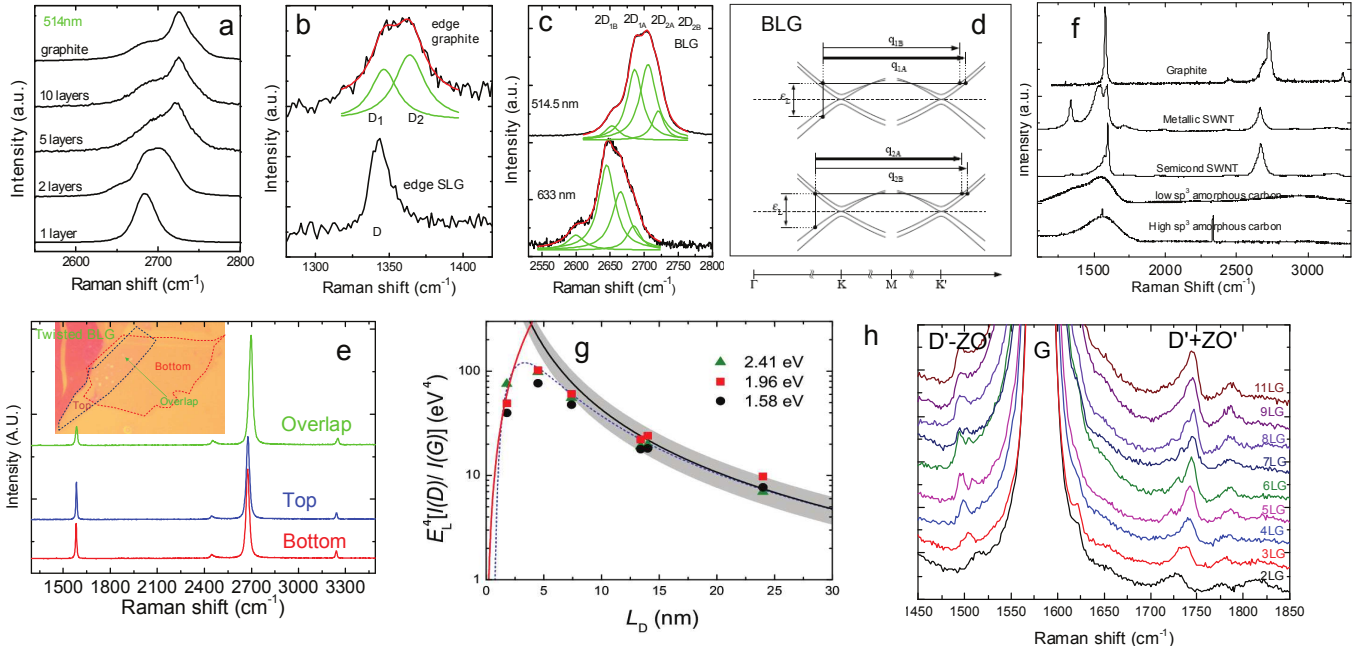}}
\caption{\textbf{Dependence of the Raman spectra on number of layers and disorder}. \textbf{(a)} 2D peak as a function of $N$ for 514nm excitation. Flakes are labeled as in Fig.1f \textbf{(b)} D peak at the edge of graphite and SLG. \textbf{(c)} 2D peak of BLG measured at 514.5 and 633nm. In \textbf{b,c} the red curves show multiple Lorentzian fits, and the green curves show the individual Lorentzian components, labeled by D with corresponding subscripts. \textbf{(d)} Phonon wave vectors corresponding to the 2D peak in BLG. \textbf{(e)} Raman spectra of two individual SLG juxtaposed to create a twisted BLG. The inset shows the sample. \textbf{(f)} Raman spectra of graphite, metallic and semiconducting nanotubes, low and high $sp^3$ amorphous carbons.\textbf{(g)}~Amorphisation trajectory -- $E_L^4[I(D)/I(G)]$ as a function of $L_D$ for different excitation energies~$E_L$ \cite{cancadoacf,Ferrari00}. The dashed blue line is obtained from the empirical model of Ref.\onlinecite{lucchese}. The solid black and red lines are $\propto{L}_D^{-2}$ and $\propto{L}_D^2$ dependences, respectively. The shaded area accounts for the upper and lower limits given by the 30\% experimental error.\textbf{(h)} Combination modes close to the G peak as a function of $N$. Flakes are labeled as in Fig.1f. The peaks$\sim$1500cm$^{-1}$ and between$\sim$1700 and 1840cm$^{-1}$ are assigned as anti-Stokes and Stokes combinations of the E$_{2g}$ LO and the B$_{2g}$ ZO' (layer-breathing mode) phonons, respectively\cite{Lui1,Maul}}
\end{figure*}

We now examine bulk graphite\cite{ACFRaman,cancado08}. In this case the electronic energy depends on the wave vector along the (001) direction, $k_\bot$. For a fixed $k_\bot$, the in-plane dispersion is similar to that of BLG, with the splitting between the branches depending on $k_\bot$. In principle, the excited phonons can have any perpendicular wave vector $q_\bot$, however, the main contribution comes from the one-dimensional van Hove singularity at $q_\bot=0$. The resulting in-plane phonon momenta $\vec{q}_\mathrm{1B}$, $\vec{q}_\mathrm{1A}$, $\vec{q}_\mathrm{2A}$, $\vec{q}_\mathrm{2B}$ depend on $k_\bot$ of the excited electron and hole. Summation over all possible $k_\bot$ gives a distribution of allowed $\vec{q}$'s, in turn associated to a Raman frequency distribution. The lower branch of the graphite electronic dispersion has weaker dependence on $k_\bot$ than the upper one. Thus, also $\vec{q}_\mathrm{2B}$, corresponding to the only process not involving the upper branch, has weaker dependence on $k_\bot$ than $\vec{q}_\mathrm{1B}$, $\vec{q}_\mathrm{1A}$, $\vec{q}_\mathrm{2A}$. As a result, $2D_\mathrm{2B}$ appears as a sharp peak, while others merge into a low-frequency shoulder.

Not all multilayers are Bernal stacked. In fact, it is possible to observe more complex 2D shapes than those described thus far, e.g., in samples grown by carbon segregation or chemical vapor deposition, as well as sometimes in bulk or exfoliated samples\cite{pellegrini,kopel}. In principle, any relative orientation and stacking of the graphene layers could be possible, and this would be reflected in a significant change of the band structure\cite{latil}. Indeed, some orientations and stackings do result in a Dirac-like spectrum\cite{latil}, thus giving a SLG-like 2D peak, even in few-layer graphene (FLG). It was noted in early studies that turbostratic graphite has a single 2D peak\cite{lespade}, but with FWHM(2D) almost double than SLG, and Pos(2D) upshifted by $\sim{2}0\:\mbox{cm}^{-1}$. Turbostratic graphite has often a D peak\cite{lespade}, thus defect induced broadening. It is possible to prepare multilayers with any relative orientation\cite{bona,bonag}. Fig.4e compares the spectra of two SLG and a BLG formed by their juxtaposition. The SLG 2D peak shape is preserved in the twisted BLG. Thus, a detailed study of the 2D peak shape can bear important information on the multilayer interactions, and the presence of Dirac electrons in FLG\cite{pellegrini,kopel}. On the other hand, especially when characterizing grown samples, care should be taken in asserting that a single-Lorentzian 2D peak is proof of SLG. This could also be consistent with turbostratic or twisted FLG, especially if broader than in exfoliated SLG. By carefully studying the shape and width of the 2D band, it is then possible to distinguish between different stackings\cite{Lui2011} and relative orientations\cite{Kim2012}. Another signature of twisted BLGs is the appearance of two one-phonon peaks, one in the region$\sim$1350-1550$cm^{-1}$, and the other around $1620\:\mbox{cm}^{-1}$. These peaks, non-dispersive with $\omega_L$, are due to phonons with wave-vectors determined by the period of the superlattice that is formed by the superposition of the two SLGs and that relaxes the fundamental selection rule $\vec{q}=0$ (Refs.\onlinecite{Gupta2010,Carozo2011}).

It is also interesting to note that single-wall nanotubes show a sharp 2D peak, similar to SLG (ref.~\onlinecite{jorio}), while double or multiple peaks can be observed in multi-wall tubes\cite{pfeiffer}. Their shape and position bears information on the interaction and relative orientation of each tube, and their diameters.

As discussed in Section~\ref{sec:spectrum}, another method to probe $N$ and stacking order in multilayer graphene is based on Raman signatures of out-of-plane vibrations\cite{Lui1,Lui2,Sato,Maul}. In SLG, the two out-of-plane modes $A_{2u}$ and $B_{2g}$ (Fig.1b) are not Raman-active. Fig.4h shows that close to the G peak there are other features whose position changes with~$N$. Of interest is the peak between 1700 and $1750\:\mbox{cm}^{-1}$ assigned as the combination of the $E_{2g}$ LO~phonon and the $B_{2g}$ ZO$'$~phonon (LBM)\cite{Lui1,Maul}. This corresponds to a double resonant process and its frequency depends on the laser excitation energy~$E_L$. Its anti-Stokes combination can also be seen in Fig.4h below the G peak. Moreover, in FLG this splits into several components, as the number of LBMs and their frequencies depend on $N$ and on the stacking order. The C peak also bears direct information on interlayer coupling, scaling with $N$ (Fig.1f) (Ref.\onlinecite{tan42}). Note that, even though the 2D peak shape reflects the change in the band structure with~$N$ (and this very change is what allows one to determine~$N$), the phonons involved are always in-plane, while the C and LBM phonons involve the relative motion of the planes, thus are a direct probe of layering.

\section{Defects and disorder}

Quantifying defects in graphene related systems, which include a large family of $sp^{2}$ carbon structures, is crucial both to gain insight in their fundamental properties, and for applications. In graphene, this is a key step towards the understanding of the limits to its ultimate mobility\cite{ni,kim,fuhrer}.

Ref.\onlinecite{Ferrari00} introduced a three stage classification of disorder, leading from graphite to amorphous carbons, that allows to simply assess all the Raman spectra of carbons: (1) Graphite to nanocrystalline graphite; (2) nanocrystalline graphite to low $sp^3$ amorphous carbon; (3) low $sp^3$ amorphous carbon to high $sp^3$ amorphous carbon. In the study of graphene, stages 1 and 2 are the most relevant and are summarized here.

In stage 1 the spectrum evolves as follows\cite{Ferrari00}: (a) a D peak appears and $I(\mathrm{D})/I(\mathrm{G})$ increases; (b) a D$'$ peak appears; (c) all peaks broaden, so that D and 2D lose their doublet structure in graphite; (d) the $\mathrm{D}+\mathrm{D}'$ peak appears; (e) at the end of stage 1, G and D$'$ are so wide that is sometimes more convenient to consider them as a single, upshifted, wide G band at $\sim1600\:\mbox{cm}^{-1}$.

In their seminal work, Tuinstra and Koenig noted that $I(\mathrm{D})/I(\mathrm{G})$ varied inversely with the crystal size,~$L_a$: $I(\mathrm{D})/I(\mathrm{G})=C(\lambda)/L_a$, where $C(514\:\mbox{nm})\sim{4.4}\:\mbox{nm}$\cite{tk,matt,knight} ($\lambda=\pi{c}/\omega_L$ is the excitation wavelength). Initially, this was understood in terms of phonon confinement: the intensity of the forbidden process would be ruled by the ``amount of lifting'' of the selection rule\cite{tk}, $\Delta q\propto 1/\Delta x$, where the coordinate uncertainty $\Delta x\approx L_a$. Now, it is understood theoretically and established experimentally that the D peak is produced only in a small region of the crystal (size $\sim{v}_F/\Omega^{TO}\sim{3-4}\:\mbox{nm}$) near a defect or an edge\cite{Cedge,Basko09,lucchese,beams}. For a nanocrystallite, $I(\mathrm{G})$ is proportional to the sample area, $\propto{L}_a^2$, while $I(\mathrm{D})$ is proportional to the overall length of the edge, which scales as~$L_a$. Thus, $I(\mathrm{D})/I(\mathrm{G})\propto1/L_a$. For a sample with rare defects, $I(\mathrm{D})$ is proportional to the total number of defects probed by the laser spot. Thus, for an average interdefect distance $L_D$, and laser spot size $L_L$, there are on average $(L_L/L_D)^2$ defects in the area probed by the laser, thus $I(\mathrm{D})\propto(L_L/L_D)^2$. On the other hand, $I(\mathrm{G})$ is proportional to the total area probed by the laser $\propto{L}_L^2$, thus $I(\mathrm{D})/I(\mathrm{G})=C''(\lambda)/L_D^2$. For very small $L_D$, one must have $C''(\lambda)/L_D^2=I(\mathrm{D})/I(\mathrm{G})=C(\lambda)/L_a$. This condition gives an estimate of $C''(514\:\mbox{nm})\sim{90}\:\mbox{nm}$. Ref.\onlinecite{lucchese} measured $I(\mathrm{D})/I(\mathrm{G})$ for irradiated SLG with known $L_D$, derived from scanning tunneling microscopy, obtaining $I(\mathrm{D})/I(\mathrm{G})\approx 145/L_D^2$ at 514nm, in excellent agreement with this simple estimate (Fig.4g).

For an increasing number of defects, the Tuinstra-Koenig relation, in either form, must eventually fail\cite{Ferrari00}. The ``molecular'' picture outlined in Appendix S4 helps understanding what happens\cite{Ferrari00}. For high disorder, $sp^2$ clusters become smaller and the rings fewer and more distorted, until they open up. As the G peak is just related to the relative motion of $sp^2$ carbons, we can assume $I(\mathrm{G})$ to be roughly constant as a function of disorder. Thus, with the loss of $sp^2$ rings, $I(\mathrm{D})$ will now decrease with respect to $I(\mathrm{G})$ and Tuinstra-Koenig relation will no longer hold. For very small $L_a$ or $L_D$, the D peak strength is proportional to the number of 6-fold rings in the laser spot\cite{Ferrari00}. Thus, in amorphous carbons the development of a D peak indicates ordering, exactly the opposite to graphene\cite{Ferrari00}. The proportionality of $I(\mathrm{D})/I(\mathrm{G})$ to the number of rings leads to a new relation for stage 2: $I(\mathrm{D})/I(\mathrm{G})=C'(\lambda)L_D^2$. Imposing continuity with the Tuinstra-Koenig relation at $L_a=3\:\mbox{nm}$ gives $C'(514\:\mbox{nm})\sim 0.55$ (Ref.\onlinecite{Ferrari00}) (Fig.4g).

The spectrum in stage 2 evolves as follows\cite{Ferrari00}: (a) Pos($G$) decreases from$\sim{1600}\:\mbox{cm}^{-1}$ towards$\sim{1510}\:\mbox{cm}^{-1}$; (b) the Tuinstra-Koenig relation fails; $I(\mathrm{D})/I(\mathrm{G})\to{0}$; (c) Pos(G) acquires a dispersion as a function of excitation energy, the bigger the stronger the disorder; (d) there are no more well defined second-order peaks, but a small modulated bump between 2300-$3200\:\mbox{cm}^{-1}$ (Fig.4f).

In disordered carbons Pos(G) increases as the excitation wavelength decreases, from IR to UV\cite{Ferrari01}. Disp(G) increases with disorder. This separates the materials into two types. In those with only $sp^2$ rings, Disp(G) saturates at$\sim{1600}\:\mbox{cm}^{-1}$, Pos(G) at the end of stage 1. In contrast, for those containing $sp^2$ chains (such as in amorphous and diamond-like carbons), Pos(G) continues to rise past $1600\:\mbox{cm}^{-1}$ and can reach $1690\:\mbox{cm}^{-1}$ for 229nm excitation\cite{Ferrari01}.

The D peak always disperses with excitation energy, however, the higher the disorder, the lower Disp(D), opposite to G (Ref.\onlinecite{Ferrari01}). Finally, FWHM(G) always increases with disorder\cite{prbcn}. Thus, combining $I(\mathrm{D})/I(\mathrm{G})$ and FWHM(G) allows one to discriminate between stages 1 or 2, since samples in stage 1 and 2 could have the same $I(\mathrm{D})/I(\mathrm{G})$, but not the same FWHM(G), being this much bigger in stage 2 (Refs.\onlinecite{prbcn,cancadoacf}).

It is important to understand what the maximum in Fig.~4g means. $I(\mathrm{D})$ increases as long as the contributions from different defects add up independently. However, if two defects are closer than the average distance an $e,h$ pair travels before scattering with a phonon, then their contributions will not sum anymore. Sec.3 estimated it as $v_F/\Omega^{TO}\sim{3}.5\:\mbox{nm}$. This agrees with the predictions of Ref.\onlinecite{Ferrari00} and the measurements in Refs.\onlinecite{lucchese,cancadoacf}.

The excitation energy dependence of the peaks areas and intensities, discussed in Sec.~\ref{sec:processes}, can be exploited to generalize the amorphisation trajectory for any visible excitation energy. Fig.4g plots $E_L^4[I(\mathrm{D})/I(\mathrm{G})]$ as a function of $L_D$, as measured in Ref.\onlinecite{cancadoacf}, where $E_L=\hbar\omega_L$ is the laser excitation energy in eV. The data with $L_D>10\:\mbox{nm}$ obtained with different laser energies collapse in the same curve. In this regime the $I(\mathrm{D})$ is proportional to the number of point defects, giving rise to $I(\mathrm{D})/I(\mathrm{G})\propto{1}/L_D^2$, as discussed above. A fit to the experimental data gives\cite{cancadoacf}:
\begin{equation}\label{eq12}
L_D^2\,{\rm(nm^{2})}=\frac{(4.3\pm1.3)\times10^{3}}{E_L^4(\mbox{eV}^4)}\left[\frac{I(\mathrm{D})}{I(\mathrm{G})}\right]^{-1}
\end{equation}
In terms of defect density $n_D^2=1/(\pi{L}_D^2)$,
$n_{\rm D}(\mbox{cm}^{-2})=10^{14}/(\pi L_{\rm D}^2)$:
\begin{equation}\label{eq14}
n_{\rm D}({\rm cm}^{-2})=(7.3\pm2.2)\times10^{9}E_L^4(\mbox{eV}^4)\,\frac{I(\mathrm{D})}{I(\mathrm{G})}
\end{equation}
For the high defect density regime (stage 2, with $L_D<3\:\mbox{nm}$), $I(\mathrm{D})$ decreases with respect to $I(\mathrm{G})$ as $L_D$ decreases, leading to $I(\mathrm{D})/I(\mathrm{G})\propto{L}_{\rm D}^2$ (Ref.\onlinecite{Ferrari00}). The red line in Fig.4g is the fit to the data with $L_{\rm D}$<3nm giving:
\begin{eqnarray}\label{eq101}
&&L_{\rm D}^{2}{\rm(nm^{2})}=(5.4\times10^{-2})\,E_L^4(\mbox{eV}^4)\,\frac{I(D)}{I(G)},\\
\label{eq102}
&&n_{\rm D}^{2}{\rm(cm^{-2})}=\frac{5.9\times10^{14}}{E_{\rm L}^4}\left[\frac{I(\mathrm{D})}{I(\mathrm{G})}\right]^{-1}
\end{eqnarray}
The proportionality factor in Eq.\ref{eq101} at 2.41eV is$\sim(0.55)^{-1}$, in excellent agreement with that proposed by Ref.\onlinecite{Ferrari00}.

Note that these relations are of course limited to Raman active defects, but one has to bear in mind that there are also non-Raman active defects. E.g. perfect zigzag edges\cite{Cancado04,Cedge}, charged impurities\cite{cinzapl,das1}, intercalants \cite{tandoping}, uniaxial and biaxial strain\cite{Mohi,halsall} do not generate a D peak. For these types of ``silent'' defects, other Raman signatures can be used. A perfect zig-zag edge does change the G peak shape\cite{Sasaki09,Cong10}, while strain, intercalants, and charged impurities have a strong influence on the G and 2D peaks\cite{cinzapl,das1,tandoping,Mohi}. In this case, the combination of Raman spectroscopy with other independent probes of the number of defects can provide a wealth of information on the nature of such defects.
\section{Effect of perturbations}
The Raman spectrum of graphene is quite sensitive to changes in many external parameters, such as strain, gate voltage, magnetic field and so on, making it a powerful and useful characterization tool. These can modify (i) the phonon properties (i.e., their energies and decay rates), or (ii) the Raman process itself. Effects of the first type can only change positions and widths, not frequency-integrated intensities, because the total spectral weight of each phonon state is conserved under perturbations. Thus, by studying peak areas and their dependence on various parameters one can extract detailed information on the Raman process, in particular on the electronic excitations, which are intermediate states. This is especially true for triple-resonant processes, among which the 2D process is the most convenient to analyze. The 2D$'$ process will be analogous to the 2D process in most situations, but the 2D peak is more intense due to Coulomb-induced EPC enhancement\cite{baskoal,Lazzeri08,BPF09}.
Appendixes S6-S10 overview the effect of various types of perturbation on the Raman spectra: electric field and doping (S6), magnetic field (S7), uniaxial and biaxial
strain (S8), temperature (S9), isotopic composition (S10).
\section{Beyond graphene}
The quest to understand the Raman spectrum of graphite started over 40 years ago\cite{tk}. This may be the longest any Raman spectrum has been studied for, and is bound to continue for years to come. The availability of monolayer graphene widened even more, if at all possible, the appeal for this technique. The study of Raman scattering in other layered materials has just begun. We note that the seminal work of Novoselov \textit{et al.}\cite{pnas} not only indicated how to produce graphene by micromechanical cleavage, but also reported layers of other materials, such as MoS$_2$. The Raman spectra of the bulk counterparts of these materials were studied many years ago, often by the same authors interested in graphite, such as Nemanich et al. for boron nitride\cite{nemanichbn}. Now that individual monolayers and multilayers are available, there is a resurgence of Raman studies on them. Many of these are polar materials, thus will behave differently from graphene as a function of $N$, as discussed in Ref.\onlinecite{arenal} for boron nitiride. Also, some of them are wide-band gap and greatly benefit from UV Raman measurements to reach resonance\cite{ReichBN}. The current interest in topological insulators, especially based on materials like BiSe, BiTe, finds in Raman spectroscopy\cite{russo} an obvious means to probe their phonons and electronic interactions down to their individual units (e.g. a quintuplet in the case of BiSe). Thus far, MoS$_2$ is one of the most studied by Raman spectroscopy. The Raman spectrum of bulk MoS$_2$ consists of two main peaks at$\sim382\:\mbox{cm}^{-1}$ and$\sim407\:\mbox{cm}^{-1}$ assigned to the $E^1_{2g}$ in-plane and $A_{1g}$ out of plane modes, respectively\cite{Verb}. The former red shifts, while the latter blue shifts with $N$ (Ref.\onlinecite{LeeH}). Moreover, they have opposite trends when going from bulk MoS$_2$ to single-layer MoS$_2$, so that their difference can be used to monitor~$N$ (Ref.\onlinecite{LeeH}). However, the trends are not fully understood and more work is needed to clarify the changes with N. Raman spectroscopy of C and LB modes is also a useful tool to probe these materials. These modes change with $N$, with different scaling for odd and even $N$ (Ref.\onlinecite{tanmos2}). The combinations of two-dimensional crystals in three-dimensional stacks could offer huge opportunities in designing the functionalities of such heterostructures\cite{novos2d,bonag} One could combine conductive, insulating, superconducting and magnetic 2d materials, tuning the performance of the resulting material\cite{novos2d}, the functionality being embedded in the design of such heterostructures\cite{novos2d}. Layered materials can be exploited for the realization of heterostructures. The interactions between different layers inside heterostructures and hybrids is expected to be weak if van der Waals forces hold them together. In this case, the vibrations of heterostructures and hybrids will consist of those from the individual building blocks. Therefore, Raman spectroscopy will be a useful tool to probe the stoichiometry of heterostructures and hybrids.
\setcounter{section}{0}
\setcounter{figure}{0}
\setcounter{equation}{0}
\renewcommand{\theequation}{S\arabic{equation}}
\renewcommand{\thefigure}{S\arabic{figure}}
\renewcommand{\thesection}{S\arabic{section}}
\section*{Appendix}
\section{Raman scattering}
Raman scattering\cite{Landsberg,Raman} is the inelastic scattering of photons by phonons (or other excitations), Fig.~S1. A photon impinging on a sample creates a time-dependent perturbation of the Hamiltonian. Due to the photon fast changing electric field, only electrons respond to the perturbation. The electronic wave-functions of the perturbed system can be written as a linear combination, with time-dependent coefficients, of all the wave-functions of the unperturbed system. The perturbation introduced by a photon of energy $\hbar\omega_L$ increases the total energy to $E_{GS}+\hbar\omega_L$, where $E_{GS}$ is the ground state energy. In general, $E_{GS}+\hbar\omega_L$ does not correspond to a stationary state, therefore the system is said to be in a virtual level. In classical language, a virtual level corresponds to a forced oscillation of the electrons with a frequency $\omega_L$. Once the photon realizes that the system has no stationary state of energy $E_{GS}+\hbar\omega_L$, it leaves this unstable situation. We can formally consider the photon as being emitted by the perturbed system, which jumps back to one of its stationary states.
\begin{figure}
\centerline{\includegraphics [width=90mm]{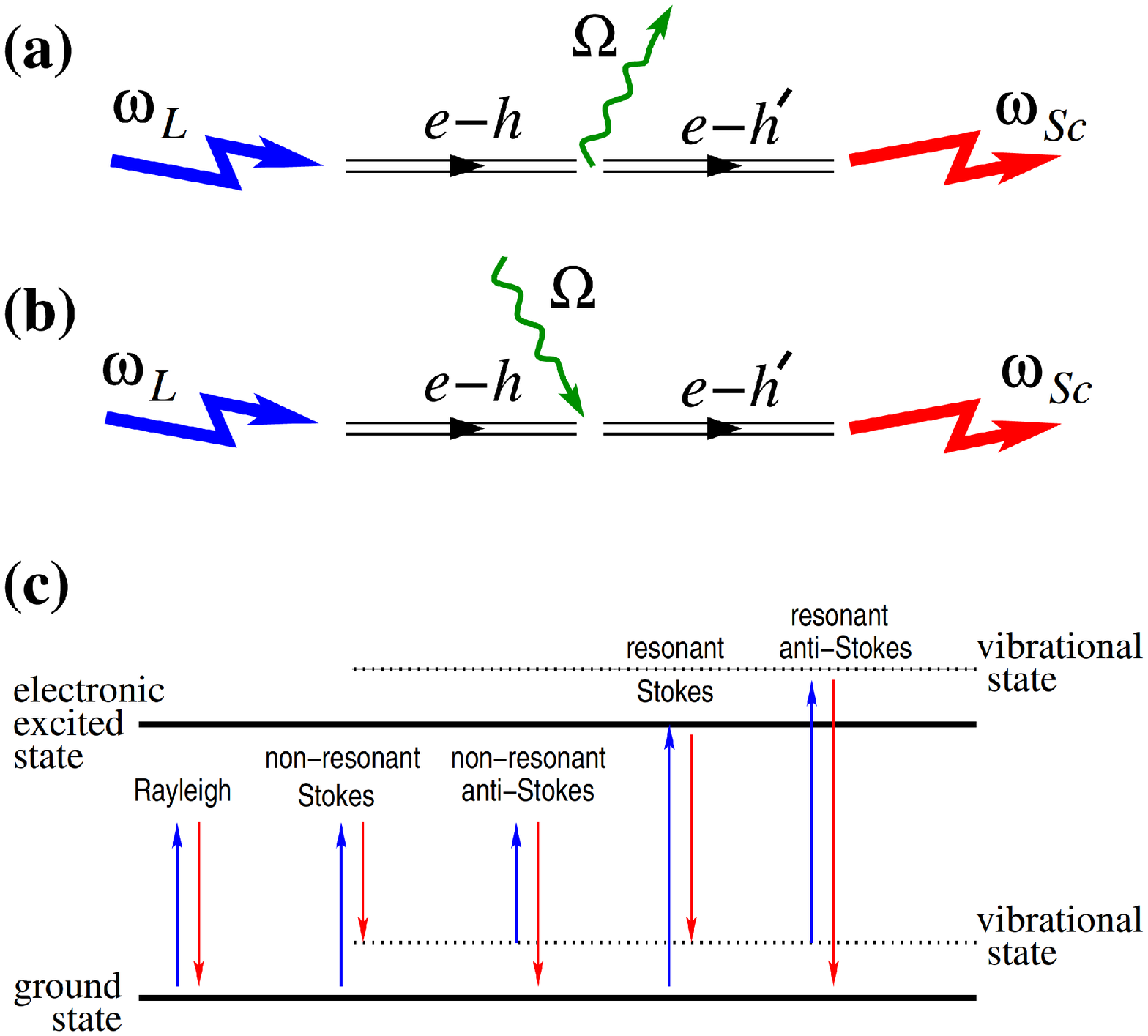}}
\caption{\textbf{Raman Scattering}. \textbf{(a)} Stokes. An incoming photon $\omega_L$ excites an electron-hole pair \textit{e-h}. The pair decays into a phonon $\Omega$ and another electron-hole pair \textit{e}-$h'$. The latter recombines, emitting a photon $\omega_{Sc}$. \textbf{(b)} Anti-Stokes. The phonon is absorbed by the electron-hole pair. \textbf{(c)} Rayleigh and Raman scattering in resonant and non resonant conditions.}
\end{figure}

\textit{Rayleigh scattering} takes place when the system returns to its initial state, and the frequency of the emitted photon remains the same as the incident one. This is also called \textit{elastic scattering}, and all that can happen to the photon is a change in its propagation direction. Still, Rayleigh or elastic scattering can give useful information\cite{tony,CasiraghiNL}. Graphene is indeed one of the most successful examples\cite{CasiraghiNL}, since elastic scattering is now commonly used to image the flakes and derive the number of layers\cite{GeimAPL,CasiraghiNL}. The presence of an appropriate spacer, such as the typical 300~nm SiO$_2$ over Si\cite{pnas}, can enhance significantly the incident field amplitude within graphene, thus its visibility\cite{GeimAPL,CasiraghiNL}. Had not been for this effect, that enabled to see SLG flakes simply using an optical microscope\cite{pnas}, the whole graphene research field may never have gained momentum.

\textit{Raman scattering} happens when, with a much lower probability than Rayleigh scattering, the photon can lose part of its energy in the interaction process, thus exiting the sample with a lower energy $\hbar\omega_{Sc}$. This corresponds to the Stokes (S) process. Since the sample has to return to a stationary state, the energy loss must correspond to a phonon energy, $\hbar\omega_L-\hbar\omega_{Sc}=\hbar\Omega$. If the incoming photon finds the sample in an excited vibrational state, and after the interaction the system returns to its ground level, the photon can leave the crystal with an increased energy $\hbar\omega_{Sc}=\hbar\omega_L+\hbar\Omega$. This corresponds to the Anti-Stokes (AS) process. Given that S is the most probable\cite{YuCardona}, the vast majority of Raman spectra in literature are S measurements plotting the intensity of the scattered light as a function of the difference between incident and scattered photon energy, the so called Raman shift. Even though the Raman shift units should be those of energy, it is historically plotted in cm$^{-1}$. These can be converted in meV using the relation $1\:\mbox{meV}=8.0655447\:\mbox{cm}^{-1}$. The S/AS ratio depends on the sample temperature\cite{YuCardona}, and is a very popular method to monitor it. However, in graphene and nanotubes, the nature of the resonant process has to be carefully considered when comparing S/AS intensities, since the resonance windows for incident and scattered photons are different\cite{oron,berciaud}. Not doing so gives wrong temperature estimations.

\textit{Non-resonant} Raman scattering is when $E_{GS}+\hbar\omega_L$ does not correspond to a stationary state, as is indeed the case for most materials. If the excitation is selected to match a specific energy level\cite{YuCardona}, then the process is \textit{resonant}, and the intensities are strongly enhanced, as a result of the greater perturbation efficiency. In a quantum mechanical description, this corresponds to a vanishing denominator in the perturbation theory expression for the transition amplitude. Resonant Raman scattering has been intensely studied since the 1960-1970s for semiconductors (see Ref.\onlinecite{Falicov} for a review).

For IR-UV light, the main scattering mechanism involves electronic excitations as intermediate states, rather than direct photon-phonon coupling, because the laser energy is large compared to the phonon energy\cite{Born}. This is even more so for carbon allotropes, since they are non-polar. Thus, the study of the Raman spectra can shed light on the behavior of electrons\cite{BPF09,Faugeras2010}, and complement transport measurements. The Raman intensities calculations in solids date back to the 1960s, with free electron-hole ($e$-$h$) pairs\cite{Loudon} and excitons\cite{Birman} as intermediate states. Graphene differs from usual semiconductors in several aspects, with important consequences. First, the linear gapless electronic dispersion implies resonances for any excitation frequency. Second (related to the first), there are no excitons in the sense of real bound states of the electron and the hole. Third, in the IR-visible range, the electronic spectrum has approximately symmetric conduction and valence bands, while for semiconductors the difference between $e$ and $h$ effective masses $m_e,m_h$ is usually of the order of $m_e,m_h$ themselves, and often $m_h\gg{m}_e$\cite{YuCardona}. Still, Raman scattering for semiconductors with equal masses was previously studied, both theoretically and experimentally. In indium halides (such as InBr,InI) $m_e=m_h$ and their Raman spectra were found to exhibit peaks up to 20th order; even-order peaks being more intense than odd-order ones~\cite{In1}. This was assigned to full resonance\cite{Lang90}, which is of direct relevance for even-order peaks in graphene.

If $\vec{k}_L$ and $\omega_L=ck_L$ are the wave vector and frequency of the incoming photon, $\vec{k}_{Sc}$ and $\omega_{Sc}=ck_{Sc}$ those of the scattered photon ($c$~being the speed of light), and $\vec{q}$ and $\Omega^\nu_\vec{q}$ those of a phonon belonging to the branch~$\nu$, then energy and momentum conservation give:
\begin{eqnarray}
&&\omega_L=\omega_{Sc} \pm \Omega_\vec{q}^\nu,\\
&&\vec{k}_L=\vec{k}_{Sc} \pm \vec{q}.
\end{eqnarray}
In the S process a phonon is created (``$+$''), in AS one is annihilated (``$-$''). Typical Raman experiments are conducted in the $1064-229$nm range, corresponding to $1.2-5.4$eV\cite{pocsik,Ferrari00,acftrans}. Since the lattice parameter, $a$, is of the order of a few {\AA} ($1.42\:\mbox{\AA}$ in graphene\cite{charlier}), $k_L,k_{Sc}\ll\pi/a$, the magnitude of a zone boundary wavevector. Then, from Eqs.(S1,S2), $q\ll\pi/a$, i.e. in first-order scattering only phonons near $\vec{\Gamma}$ ($\textbf{q}\approx\textbf{0}$) are measured. This is sometimes referred to as the \textit{fundamental Raman selection rule}.

The emission of two phonons with opposite wavevectors can \textit{always} satisfy the fundamental selection rule: $\vec{q}+(-\vec{q})=0$. Since each individual $\vec{q}$ \textit{a priori} can be arbitrary, all phonons may contribute to a multi-phonon process. Thus, in principle the multiphonon Raman spectrum will reproduce the main features of the phonon density of states (PDOS), and can be used to derive important information on the phonon branches, as done, e.g. in hexagonal boron nitride, h-BN\cite{ReichBN}. Graphene is yet again special: only very few intense features are seen, not corresponding to the PDOS maxima\cite{ACFRaman,piscanec}, due to the peculiar nature of the resonant process and the electron-phonon and electron-electron interactions\cite{ACFRaman,piscanec,BPF09}. h-BN, albeit with an hexagonal lattice as graphene, is a wide gap semiconductor and, unless Raman spectroscopy is performed in the deep UV\cite{ReichBN,arenal}, is not resonant.

In general, Raman scattering can be described by perturbation theory\cite{YuCardona}. For an $n$-phonon process we have an $(n+2)$ order matrix element:
\begin{widetext}
\begin{equation}\label{Ramanmatrixelement=}
\mathcal{M}=\sum_{s_0,\ldots,s_n}\frac{\langle{f}|\hat{H}^\mathrm{em}|s_n\rangle \langle{s_n}|\hat{H}^\mathrm{ph}|s_{n-1}\rangle \ldots\langle{s_1}|\hat{H}^\mathrm{ph}|s_0\rangle\langle{s}_0|\hat{H}^\mathrm{em}|i\rangle} {(\hbar\win-E_n+i\Gamma_n/2)\ldots(\hbar\win-E_1+i\Gamma_1/2)(\hbar\win-E_0+i\Gamma_0/2)}
\end{equation}
\end{widetext}
with $|i\rangle$ the initial state (incident photon with frequency $\win$, polarized along a unit vector~$\ein$), $|f\rangle$ the final state (scattered photon with frequency $\omega_{Sc}$ and polarization $\eout$, and $n$~phonons left in the crystal), while~$s_k$, $k=0,\ldots,n$, label the intermediate states where no photons are present, but an $e-h$ pair is created and $k$~phonons emitted. $E_k$ and $\Gamma_k/\hbar$ are the energies and decay rates of these intermediate states. $\hat{H}^\mathrm{em}$~and~$\hat{H}^\mathrm{ph}$ are the Hamiltonians describing the interaction of electrons with the electromagnetic field and with phonons. In practice, for $n>1$, the enumeration of all possible intermediate states is cumbersome, so that a diagrammatic scattering formalism is useful for a systematic analysis of all relevant terms. According to the number of vanishing denominators in Eq.~(S3), the process can be classified as double-resonant or triple-resonant. Higher orders are also possible in multiphonon processes.

Each term in the sum of Eq.~(\ref{Ramanmatrixelement=}) can be viewed as the complex amplitude of the corresponding elementary process with given intermediate states. These amplitudes may add up in phase or out of phase, which would correspond to constructive or destructive quantum interference. For non-interacting electrons in a defect-free crystal, the summation over the intermediate states in Eq.~(\ref{Ramanmatrixelement=}) is reduced to an integration over the electronic momentum. In the presence of defects, the momentum is not a good quantum number for electronic wave functions. Still, if the density of defects is not very high, their effect on the wave functions can be treated perturbatively. This corresponds to introducing more intermediate states, as well as matrix elements of electron-defect scattering, $\langle{s}_{i+1}|\hat{H}_{def}|s_i\rangle$, in Eq.~(\ref{Ramanmatrixelement=}), and again integrating over the electronic momentum. This procedure treats electron scattering on defects analogously to scattering on phonons, so that a defect can be viewed as a phonon with zero frequency.

Given $\mathcal{M}$, one should sum its square over the phonon wavevectors, either with fixed total energy, to obtain the frequency-resolved intensity $I(\omega)$, or over all energies, to get the frequency-integrated peak intensity~$A$:
\begin{eqnarray}
&&I(\omega_{Sc})\propto\int|\mathcal{M}|^2 \delta(\Omega^{\nu_1}_{\vec{q}_1}+\ldots+\Omega^{\nu_n}_{\vec{q}_n}-\omega_{Sc})\, d^2\vec{q}_1\ldots{d}^2\vec{q}_n,\nonumber\\ &&\\
&&A=\int\limits_{\omega_{Sc}\in\mathrm{peak}}{I}(\omega_{Sc})\,d\omega_{Sc} \propto\int|\mathcal{M}|^2d^2\vec{q}_1\ldots{d}^2\vec{q}_n,
\end{eqnarray}
where $\vec{q}_1,\ldots,\vec{q}_n$ are the wave vectors of the emitted phonons, $\nu_1,\ldots,\nu_n$ their branch labels, and $\Omega^{\nu_1}_{\vec{q}_1},\ldots,\Omega^{\nu_n}_{\vec{q}_n}$ are their frequencies.
The frequency-integrated intensity is more robust with respect to various perturbations of the phonon states. Indeed, for dispersionless undamped phonons, $\Omega^\nu$, the shape of the $n$-phonon peak is $\propto\delta(\omega_{Sc}-n\Omega^\nu)$ with zero width, infinite height, but well-defined area. If the phonons are allowed to decay (e.g, into other phonons due to anharmonicity\cite{Bonini2007} or into $e$-$h$ pairs due to EPC\cite{lazzeri}), the $\delta$-peak is broadened into a Lorentzian, but the area is preserved, as the total number of phonon states cannot be changed by such perturbations. If phonons have a weak dispersion, then states with different momenta contribute at slightly different frequencies. This may result in an overall shift and a non-trivial peak shape, but frequency integration across the peak means counting all phonon states, as in the dispersionless case. Thus, the frequency-integrated intensity is preserved as long as $\mathcal{M}$ is not changed significantly by the perturbation. The latter holds when the perturbation is smaller than the energy scale determining~$\mathcal{M}$. This is usually dominated by electronic broadening, and often larger. Converting energy into time by the uncertainty principle, if the process is faster than phonon decay, the total number of photons emitted within the given peak (i.~e., integrated over frequency across the peak), is not affected by phonon decay, although their spectral distribution can be.

Raman spectroscopy can also probe scattering of photons by electronic excitations. In pristine graphene electronic excitations have a continuous structureless spectrum\cite{wallace}, which does not lead to any sharp features. However, in a strong magnetic field, $B$, when the electronic spectrum consists of discrete Landau levels, the electronic inter-Landau-level excitations give rise to sharp $B$-dependent peaks in the Raman spectrum\cite{Kashuba2009,Kashuba2012,Faugeras2011,smi11}.
\section{Interference-Enhanced Raman Scattering}
In general the Raman intensity depends on the square of the incident field amplitude\cite{YuCardona}. Thus it can also be enhanced by a proper choice of substrate and spacer, resulting in the so called Interference Enhanced Raman Scattering (IERS). This is a common occurrence in graphene\cite{yoonIERS,singa}, and the reasons are analogous to those enabling its visibility on a substrate\cite{yoonIERS,singa}. Due to interference, the enhancement varies as a function of both excitation and emission wavelengths, being in principle different for different Raman peaks. The transfer matrix method can be used to evaluate the effect of substrate interference and sample absorption\cite{sers}. First, the incident amplitude $E(x)$ in the sample is evaluated, as a function of sample thickness $x$. The Raman absorption at depth $x$ is proportional to $|E(x)|^2$. Next the emissivity $E_R(x)$ from depth $x$ at the corresponding Stokes-shifted frequency is calculated\cite{sers}. The Raman intensity is then proportional to $\int _{x=0} ^{d} |E(x)|^2 |E_R(x)|^2 dx$. This shows that changes with respect to an ``intrinsic'' $I(\mathrm{2D})/I(\mathrm{G})$ are small for a SLG on 300~nm SiO$_2$ measured at 514~nm. Thus it is safe to compare this ratio amongst different samples and assign its variation to doping or other external perturbations. However, different substrates or wavelengths, may result in IERS changes of $I(\mathrm{2D})/I(\mathrm{G})$ even for the same sample. On the other hand, an optimal interference-enhancing substrate could be designed, with uniform enhancement across different Raman bands (so that their ratios do not change), at the same time ensuring high optical visibility.
\section{Surface-Enhanced Raman Scattering}
Surface enhanced Raman scattering (SERS) exploits surface plasmons induced by the incident field in metallic nanostructures, to increase the intensity \cite{kneipp06}. In principle, even a single metallic nanostructure, e.g., a nanotip, can induce SERS, giving rise to the so-called tip-enhanced Raman scattering (TERS)\cite{achim}. The key feature of TERS is its capability of optical sensing with high spatial resolution beyond the light diffraction limits\cite{achim}.

Graphene offers an ideal model system to study SERS. Ref.~\onlinecite{sers} did SERS by depositing patterned particles of different sizes and spacing on SLG. Taking into account the surface plasmon resonance near-fields, the Raman enhancement scales with particle cross section, fourth power of Mie enhancement, and inversely with the tenth power of the separation between particle centre and graphene\cite{sers}, pointing to thin nanodisks to achieve the highest SERS for 2d systems like graphene\cite{sers}.
\section{Raman scattering in Graphite and Graphene: History and Nomenclature}
Fig.1e in the main text shows the Raman spectrum of pristine and defected graphene, while Fig.4f of the main text compares the Raman spectra of graphite, nanotubes and amorphous carbons.

It is very instructive to summarize the main steps in the historical development of the understanding of the graphite Raman spectrum. This was first reported in 1970 in the seminal work of Tuinstra and Koenig (TK)\cite{tk}. They assigned the mode at $\sim{1580}$~cm$^{-1}$ to the high frequency $E_{2g}$ Raman allowed optical phonon. They also measured in defected and nanocrystalline graphite a second peak at $\sim$1350~cm$^{-1}$. They assigned it to an $A_{1g}$ breathing mode at \textbf{K}, activated by the relaxation of the Raman fundamental selection rule\cite{tk}. This was done following both the ``molecular'' and ``solid state'' routes.

The assignment is straightforward considering carbons as big molecules\cite{mapelli}. Similar bands are detected in all poly-aromatic hydrocarbons\cite{castiglioni}. The higher frequency band is due to bond stretching of $sp^2$ pairs in both rings and chains, while the lower frequency one is due breathing modes of $sp^2$ atoms in rings\cite{tk,Ferrari00,castiglioni} (Fig.1c of the main text). Thus, in absence of rings, the latter would be absent, while the former is present in any carbon materials, ranging from carbon chains, to hard amorphous carbons\cite{acftrans} (Fig.4f of the main text). Except for UV excitation, the Raman spectra of carbons are dominated by the $sp^2$ sites, because visible light resonates with the $\pi$ states, the cross section for graphite and graphene at 514~nm being$\sim$55 times higher than diamond\cite{wada}. Only for diamond, or samples with a significant fraction of diamond phase, the diamond $sp^3$ peak at 1332~cm$^{-1}$ is seen\cite{diaacf}. In amorphous carbons, the C$-$C $sp^3$ vibrations can be seen for UV excitation at $\sim$1060cm$^{-1}$ (T peak, from Tetrahedral)\cite{Ferrari00}.

The ``solid-state'' interpretation has been debated for the past 40 years.

Tuinstra and Koenig tried to combine the ``molecular'' and ``solid-state'' approaches as follows. First they considered graphite nanocrystals as big aromtic molecules, and noted that the only new Raman active modes would have $A_{1g}$ symmetry (Fig.1b of the main text). Then they looked at the graphene lattice and noted that the only BZ points with high enough symmetry to give an $A_{1g}$ mode were $\vec{K}$ and $\vec{K}'$. They were left with a problem: the fundamental Raman selection rule forbids $\bf{q\neq{0}}$ phonons. Since they observed the $\sim$1350cm$^{-1}$ band to increase with decreasing crystal size, they assumed that phonon confinement in ever smaller nanocrystals would progressively lift the selection rule. From the uncertainty principle $\Delta q\Delta x\sim\hbar$, thus $\Delta q\propto 1/\Delta x$, and the smaller the crystal size $\Delta x$, the larger $\Delta q$. However, this picture had a flaw. To activate BZ boundary $A_{1g}$ phonons requires $\Delta x\sim$ lattice spacing, in disagreement with the observation of the $\sim$1350~cm$^{-1}$ band in crystals as large as $\sim$100~nm\cite{tk}. Also, confinement could not explain why the $A_{1g}$ mode is more intense than others closer to $\bf{\Gamma}$ (all these modes could in principle be activated, as happens in other nanomaterials\cite{sinw}).

Tuinstra and Koenig did not give any names to these Raman peaks. The first nomenclature was proposed by Vidano and Fishbach in 1977\cite{vidano1}. Since they observed strong lines at $\sim$1580 and $\sim$2700~cm$^{-1}$ in pristine graphite, while other bands at $\sim$~1350 and $\sim$1620~cm$^{-1}$ only appeared in defected graphite, they called the former G, G$'$ (from Graphite) and the latter D, D$'$ (from Disorder). Nemanich and Solin detected a sharp band at $\sim$3250~cm$^{-1}$ in pristine graphite\cite{nema2}, as well as a weaker one at $\sim$2450~cm$^{-1}$\cite{nema3}. They also noted a further peak at$\sim$2950~cm$^{-1}$ in defected samples\cite{nema3}, later named D$''$ by Vidano et al.\cite{vidano2}. In 1979 Nemanich and Solin, by polarization dependent measurements, determined that all peaks between 2300 and 3250~cm$^{-1}$ in pristine graphite are overtones. In particular, they indicated that the $\sim$2450~cm$^{-1}$ peak was also an overtone. They also noted that, with defects, combinations of phonons with different wavevectors become allowed, since the requirement to have opposite wavevector is progressively lifted. They thus assigned the$\sim$2950~cm$^{-1}$ band as $\mathrm{D}+\mathrm{D}'$, rather than $\mathrm{D}+\mathrm{G}$, due to the phonon density of states (PDOS) maximum at$\sim$1620~cm$^{-1}$ for the LO branch (Fig.~1d of the main text). In 1981 Vidano \textit{et al.} studied the excitation energy dependence, and confirmed G$'$ to be the D overtone, and the $\sim$3250~cm$^{-1}$ peak the D$'$ overtone, since these shifted at twice the rate of their fundamentals\cite{vidano2}. They stressed those bands behaved differently from G, that did not move with excitation energy. They also noted the energy dependence of the $\sim$2950~cm$^{-1}$ peak was consistent with $\mathrm{D}+\mathrm{D}'$, or $\mathrm{D}+\mathrm{G}$.

Thus, by 1981 it was clear that, while the Raman-allowed first-order G peak did not shift with excitation energy, the ``defect-related'' bands D, D$'$, their overtones and combinations did. This ruled out their assignment to PDOS maxima, activated by confinement, since a PDOS maximum cannot change as a function of excitation energy, being an intrinsic material property. The symmetry and phonon branches that originated D and D$'$ were also known. The next step would have been to figure out the reason for this shift.

However, subsequent works claimed D derived from a PDOS maximum around \textbf{M} or \textbf{K}\cite{jishi}, even though phonons at \textbf{M} do not have the required symmetry, PDOS maxima are inconsistent with the D and G$'$ dispersion, and the dimension of the crystals in Ref.~\onlinecite{tk} was too big to activate zone boundary phonons by confinement only. These issues and discrepancies remained unresolved, and even ignored, for the following 16 years. A similar fate happened for other main peaks. E.g., although it was clear the$\sim$3250~cm$^{-1}$ band is not the G second order, a large number of papers (to date) still call it 2G.

In 1998 Pocsik et al.\cite{pocsik} repeated the experiments of Vidano \textit{et al.}\cite{vidano2} over a much larger excitation energy range, as shown in Fig.~1g of the main text, and, not surprisingly, found the same results. To try and explain the excitation energy dependence they proposed a ``new'' resonant process, whereby a strong enhancement of the Raman cross-section would happen for a phonon of wavevector $\vec{q}$, when this equals the wavevector $\vec{k}$ of the electronic transition excited by the incident photon (the so-called $\vec{k}$=$\vec{q}$ ``quasi-selection rule''\cite{Ferrari00}). However, the physical reason for this ``quasi-selection rule'' was unclear (it does not exist in Raman scattering) and did not yet explain why, amongst all phonons satisfying it, only those on one particular branch would be seen. Given the experimental $\mathrm{Disp}(\mathrm{D})\sim{50}$~cm$^{-1}$/eV,\cite{pocsik} only a phonon branch upshifting from \textbf{K} would satisfy the ``quasi-selection'' rule, since the linear electron dispersions select larger~$k$ with increasing excitation energy, thus larger~$q$. However, the TO branch, corresponding to the $A_{1g}$ phonon at \textbf{K}, had the opposite behavior in the most popular calculations at the time\cite{jishi}. Pocsik et al. thus identified the LO branch, with E symmetry at \textbf{K}, as responsible for the D peak, in contrast with TK. In 1999 Ref.~\onlinecite{matt} repeated the experiments of Ref.~\onlinecite{pocsik}, and reached the same conclusions\cite{matt}. Thus, 30 years after Ref.~\onlinecite{tk}, we were back to square one. A selection rule had to be ``invented'' to explain the D peak, and it was assigned to a different branch and symmetry, in contrast with the ``molecular'' view\cite{mapelli}. There was also no convincing explanation why a zone boundary phonon would be active.

In 2000, Thomsen and Reich suggested double resonance as the activation mechanism\cite{thomsen}: i) the laser excites an \textit{e-h} pair with wavevector \textbf{k} defined by resonance with the $\pi,\pi^*$ bands; ii) this is followed by electron-phonon scattering with exchanged momentum \textbf{q} near \textbf{K}; iii) defect back-scattering of the electron to the initial \textbf{k}; iv) \textit{e-h} recombination. This process allows for an exchange of a large phonon momentum, while satisfying energy conservation at any step, thus the fundamental Raman selection rule. A defect is also needed, consistent with the observations of TK. Notably, double resonance was first proposed by Baranov \textit{et al.} in 1987\cite{baranov}, who called it ``double coupled resonance'', but somehow went unnoticed, until Ref.\onlinecite{thomsen} was out. We discussed in the main text the important ramifications of this mechanism, and the current understanding of the processes responsible for the various Raman peaks, as well as outlining the still open issues.

Besides the activation mechanism, the phonons around \textbf{K} are crucial for the correct D assignment, since $\mathrm{Disp}(\mathrm{D})$ depends on the precise shape of these branches. Graphene has three branches around $\vec{K}$ which could in principle contribute (Fig.~1d of the main text). Following the suggestion of Pocsik et al.\cite{pocsik} most authors\cite{pocsik,matt,thomsen,gruneis,saito} assigned D to the LO branch stemming from the doubly degenerate $\sim$1200~cm$^{-1}$ $E$ mode at \textbf{K}, until 2004, when Ref.~\onlinecite{piscanec} finally demonstrated the D phonons belonged to the TO branch starting from the $A_{1g}$ mode at \textbf{K}. Indeed, this branch has the largest EPC amongst \textbf{K} phonons\cite{lazzeri,piscanec} and is linearly dispersive close to \textbf{K}, (Fig.~1d of the main text). A Kohn anomaly at \textbf{K}\cite{piscanec} is the physical origin of this dispersion, in quantitative agreement with the measured $\mathrm{Disp}(\mathrm{D})$, as shown in Fig.~1g of the main text.

In general, atomic vibrations are partially screened by electronic states. In a metal this screening can change rapidly for vibrations associated to certain BZ points, entirely determined by the shape of the Fermi surface. The consequent anomalous behavior of the phonon dispersion is called Kohn anomaly (KA)\cite{kohn}. KA may occur only for wavevectors \textbf{q} such that there are two electronic states $\vec{k}_1$ and  $\vec{k}_2=\vec{k}_1 + \vec{q}$ both on the Fermi surface\cite{kohn}. In graphene, the gap between occupied and empty states is zero at $\mathbf{K,K'}$ (Fig.~1a of the main text). Since $\vec{K}'=2\vec{K}$ (up to a reciprocal lattice vector), these are connected by the vector \textbf{K} (Fig.~1a of the main text). Thus, KA can occur for $\bf{q =\Gamma}$ or $\bf{q=K}$\cite{piscanec}. Ref.~\onlinecite{piscanec} demonstrated that graphene has two significant KA's for the $\bf{\Gamma}$-$E_{2g}$ and $\bf{K}$-$A_{1g}$ modes (Figs.~1a,d of the main text). It is thus impossible to derive the precise shape of the phonon branches at $\bf{\Gamma}$ and $\textbf{K}$ by approaches based on a finite number of force constants, as often done\cite{mapelli,gruneis,saito,dubay,kresse,pavone}. These results have also implications for nanotubes. Due to their reduced dimensionality, metallic tubes display much stronger KA than graphene, and folded graphene does not reproduce their phonon dispersions\cite{piscaprb,lazzeri}. The presence of KA explains the difference in the Raman spectra of semiconducting and metallic tubes\cite{piscaprb,lazzeri}.

The EPCs and phonons calculations of Ref.~\onlinecite{piscanec} were confirmed close to $\bf \Gamma$ by inelastic X-ray scattering\cite{janina,gruneis2}, and by the measurement of $\mathrm{FWHM}(\mathrm{G})$ in graphite, graphene and nanotubes\cite{lazzeri,piscaprb,ACFRaman,pisana}, once anharmonic effects are taken into account\cite{Bonini2007,ACFRaman,pisana}. A further EPC renomalization and phonon softening happens at \textbf{K} due to electron correlations\cite{baskoal,BPF09,gruneis}. Note that the $A_{1g}$ mode exactly at \textbf{K} has zero EPC for the Raman process, thus only TO phonons away from \textbf{K}, even if close to it, contribute to the D~peak\cite{piscanec,janina2}. Note as well that h-BN, even if with the same hexagonal lattice as SLG, does not have KAs, being a wide band-gap semiconductor\cite{ReichBN}.

The band at $\sim$2450~cm$^{-1}$ in Fig.~1e of the main text was first reported by Ref.~\onlinecite{nema3} in graphite, and suggested to be an overtone\cite{nema3}. Ref.~\onlinecite{nema3} remarked that, since the product of any representation with itself always contains the identity, overtones are expected to contain a contribution with polarization characteristics stronger in configurations which measure the diagonal components of the Raman tensor. They then measured the cross polarized spectrum of graphite and observed that all the high energy modes, i.e. those at $\sim$2450~cm$^{-1}$, the 2D and 2D$'$, behaved in the same way, being much stronger than the G peak compared to parallel polarization measurements. Refs.~\onlinecite{tan1,tan2,tan3} showed that this peak red-shifts with excitation energy, unlike the 2D and 2D$'$ blue-shift\cite{pocsik,vidano2}. Many alternative assignments have been put forward for this band over the years. Ref.~\onlinecite{Ferrari00} suggested a contribution from the LA branch around the BZ edge. Refs\cite{kava,tan1,tan2,tan3} interpreted it as a combination of D and a phonon belonging to the LA branch, seen at $\sim$1100~cm$^{-1}$ for visible excitation in defected samples, and called D$''$ peak (not to be confused with the old D$''$ nomenclature\cite{vidano2}, which referred to what is now know to be the D+D' peak) . Ref.~\onlinecite{shimada} assigned it as the non-dispersive overtone of the LO branch exactly at \textbf{K}, and Ref.~\onlinecite{mafra} as a combination of LA and LO phonons. On one hand combinations of phonons with different energies do not correspond to a fully resonant process\cite{baskoal,BPF09}, and would be expected to be more dominant in the presence of defects, but this band, like 2D and 2D$'$, is present in defect-free samples (Fig.~1e of the main text) while overtones exactly at \textbf{K} are non dispersive, in contrast with experiments\cite{tan1,tan2,tan3}. On the other hand, the phonons producing the D and D$''$ peaks have the same wave vector, and their energy difference is of the order of the electron scattering rate. Thus both 2D$''$ and $\mathrm{D}+\mathrm{D}''$ phonons could in principle contribute to this band, with $\mathrm{D}+\mathrm{D}''$ closer to the experimental data\cite{Ferone,Venezuela,maul2}, we thus assign it to $\mathrm{D}+\mathrm{D}''$.

To summarize, the current understanding is that the D peak is due to TO phonons around \textbf{K}\cite{tk,Ferrari00}, is active by double resonance\cite{thomsen,baranov} and is strongly dispersive with excitation energy due to the KA at \textbf{K}\cite{piscanec}. Since the so-called G$'$ band is in fact the D overtone, and has nothing to do with G, neither in terms of symmetry, nor in terms of phonon branch, nor in terms of Raman process, we renamed it 2D\cite{ACFRaman}. Consequently the $\sim$3250~cm$^{-1}$ band is 2D$'$, and the $\sim$2950~cm$^{-1}$ band is $\mathrm{D}+\mathrm{D}'$. Note that, due to resonance, in graphene (and nanotubes), it is easy to measure multiphonon peaks up to the 6th order\cite{tan1,tan2,tan3,heinz}. Our nomenclature allows to simply assign all these bands: 4D, 6D, 4D$'$, etc. With the old names, the overtones and combinations would be confusing. It was also proposed to call D$^*$ the G$'$ band, and G$^*$ the 2D$'$, using ``$*$'' to indicate second order\cite{reich1}. However, besides the problem of naming the multiphonon processes, G$^*$ is confusing since it would imply, for consistency, it being the G peak overtone, while it is in fact 2D$'$. G$^*$ is also often used to indicate the 2450~cm$^{-1}$ band\cite{mafra}, however, again, this has nothing to do with G, and this name conflicts with the use of G$^*$ for 2D$'$. Perhaps G$'$ would be a more appropriate name for D$'$, since this arises from a resonant process on the same branch giving rise to the G~peak. However, it is also true that D$'$ requires a defect and that its resonant Raman process has much in common with D, and nothing to do with G\cite{NJP}. The only drawback in calling 2D the $\sim$2700~cm$^{-1}$ band is that 2D is often used to mean ``two-dimensional''. We believe this issue to be minor, since it is hard to confuse 2D, used for ``two-dimensional'', with the $\sim$2700~cm$^{-1}$ peak. Furthermore, the nature of the 2D peak is so much related to the two dimensionality of graphene, to be not so unwarranted both share an acronym. Fig.~1e of the main text summarizes our nomenclature, when applied to defect-free, or defected graphene.
\section{Polarization dependence}
We now consider polarized excitation and detection, i. e. when the polarization vectors of incident and/or scattered light, $\vec{e}^\mathrm{in}$, $\vec{e}^\mathrm{out}$, are fixed (Fig.~S2).

The $I(\mathrm{G})$ polarization dependence can be derived from symmetry\cite{Basko08,Basko08err}. Its matrix element, described by the lattice displacement $\vec{u}$ [see Fig.~S3d for the geometry], is given by:
\begin{equation}\label{MGpolar=}
\mathcal{M}_G\propto (e^\mathrm{in}_xe^\mathrm{out}_y+e^\mathrm{in}_ye^\mathrm{out}_x)u_x +(e^\mathrm{in}_xe^\mathrm{out}_x-e^\mathrm{in}_ye^\mathrm{out}_y)u_y,
\end{equation}
where the $x$~direction is chosen to be perpendicular to the C$-$C bond.
As long as the phonon frequency does not depend on the direction of $\vec{u}$ (i.~e., the two optical modes are degenerate), $I(\mathrm{G})\propto|\mathcal{M}_\mathrm{G}|^2$, does not depend on the directions of $\vec{e}^\mathrm{in},\vec{e}^\mathrm{out}$, as shown experimentally in Fig.~S2b.
\begin{figure*}
\centerline{\includegraphics [width=150mm]{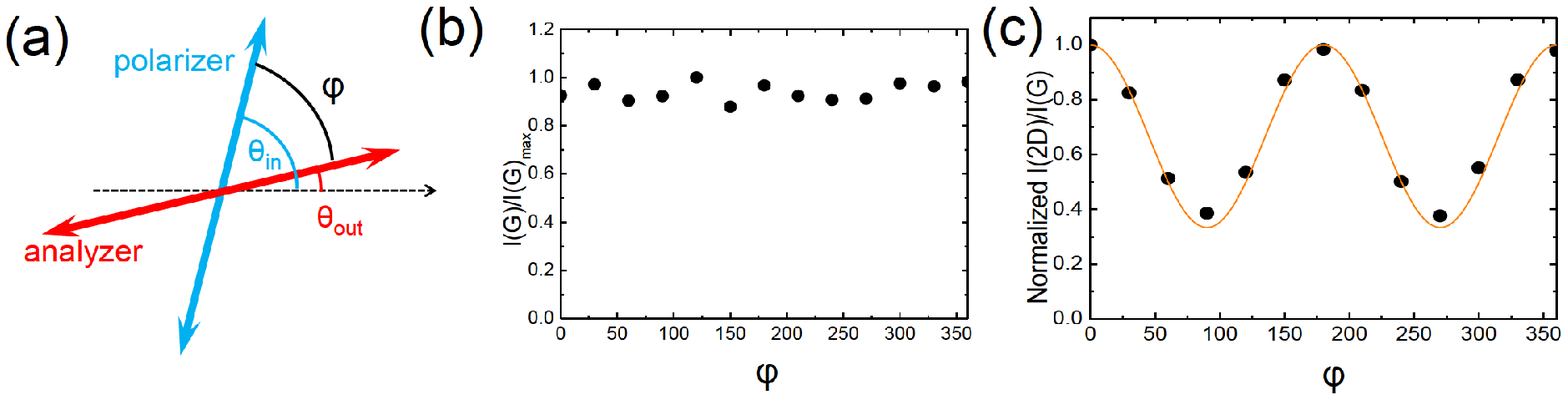}}
\caption{
\textbf{Polarization}. \textbf{(a)} Geometry for polarized measurements. Normalized \textbf{(b)} $I(\mathrm{G})$ and \textbf{(c)} $I(\mathrm{2D})/I(\mathrm{G})$, respectively, as a function of angle between polarizer and analyzer.}
\end{figure*}

The $I(2\mathrm{D})$ and $I(2\mathrm{D}')$ polarization dependence~\cite{Yoon,Basko08,Venezuela,Popov2012} can be derived from the real-space Raman picture. The interband electron-photon matrix element for photon absorption/emission is $\propto[\vec{e}^\mathrm{in/out}\times\vec{k}]$, where $\vec{k}$ is the electron momentum counted from the Dirac point. Since both $e$ and $h$ momenta are parallel to the phonon momentum $\vec{q}$ (counted from $\textbf{K}$ or $\bf{\Gamma}$ for $2\mathrm{D}$, or $2\mathrm{D}'$), the matrix element for the Raman process with emission of two phonons with momenta $\vec{q}$ and $-\vec{q}$ is $\mathcal{M}_{\vec{q}}\propto[\vec{e}^\mathrm{in}\times\vec{q}]$. The intensity is given by the integration of $|\mathcal{M}_{\vec{q}}|^2$ over the directions of~$\vec{q}$:
\begin{equation}\label{I2Dpolar=}
I(2\mathrm{D},2\mathrm{D}')\propto|\vec{e}^\mathrm{in}|^2|\vec{e}^\mathrm{out}|^2 +2(\vec{e}^\mathrm{in}\cdot\vec{e}^\mathrm{out})^2
\end{equation}
Thus, the intensity would depend on the relative orientation of $\vec{e}^\mathrm{in}$ and $\vec{e}^\mathrm{out}$, being largest when parallel and smallest (by a factor $\approx{3}$) when perpendicular, and not on their orientation with respect to the crystal. Fig.S2b,c plot the polarization dependence of $I(2\mathrm{D})$ and $I(2\mathrm{D})/I(\mathrm{G})$ for SLG, in agreement with Eq.~(\ref{I2Dpolar=}). Note that Eq.~(\ref{I2Dpolar=}) is not sensitive to the weights with which different $\vec{q}$ directions contribute. Since for each $\vec{q}$ direction there are other two, oriented at $\pm{2}\pi/3$, contributing in the same way, as required by symmetry, this is already sufficient to give Eq.~(\ref{I2Dpolar=}). The objective numerical aperture should also be taken into account when comparing with experiments.

For D and D$'$ peaks, the real-space picture is analogous to 2D, 2D$'$. Thus one can expect the same 1/3 depolarization ratio. However, this should be taken with caution, as different types of defects can scatter electrons differently, which can significantly modify the picture. E.~g., edges are defects, but give a different behavior, as discussed in Section~4 of the main text. For $\mathrm{D}+\mathrm{D}'$ the backscattering condition is absent, so the polarization memory should be weaker.

Thus far we considered both polarizer and analyzer present. The signal for unpolarized excitation is the sum of two signals for two orthogonal polarizations. The same for detection. Thus, when only one is present, there will be no dependence on polarizer/analyzer orientation.
\section{Electric Field and Doping}
An important consequence of triple resonance is the sensitivity of $A(2\mathrm{D})$ to the electronic inelastic scattering rate $2\gamma/\hbar$: $A(2\mathrm{D})\propto{1}/\gamma^2$, according to Ref.\onlinecite{Basko08} (see also Ref.\onlinecite{Venezuela}). $\gamma$ has contributions from electron-phonon and electron-electron scattering, the latter increasing with carrier concentration (or Fermi energy, $E_F$): $\gamma=\gamma_\mathrm{e-ph}+\gamma_\mathrm{e-e}$. For weak doping, $|E_F|\ll\hbar\omega_L/2$, $\gamma_\mathrm{e-e}=f|E_F|$, with the coefficient $f\sim0.5-1$ determined by the dielectric environment\cite{BPF09}. Thus, $A(2\mathrm{D})$ decreases as $E_F$ moves away from the Dirac point\cite{das1,das2}. From the measured dependence, the two contributions to $\gamma$ could be separated, giving an estimate for $\gamma_\mathrm{e-ph}\sim{20}-30\:\mbox{meV}$ at energies $\sim$1~eV\cite{BPF09}. Here it is important that the 2D peak is mostly contributed by electronic states near the $\mathbf{K}-\mathbf{M}$ direction, as discussed in the main text. Along this direction, the interband contribution to $\gamma_{e-e}$ is suppressed by trigonal warping\cite{Golub2011,Basko2013}.

Several effects originate from the fact that phonon frequencies and decay rates have a contribution due to interaction with $\pi$ electrons. This determines the phonon slopes near $\bf{\Gamma}$ and $\bf{K}$, via KAs\cite{piscanec}. This also gives a dependence of phonon frequencies and decay rates on doping\cite{Ando06,pisana,lazzeriM} and on applied magnetic field\cite{Ando07,falko}. Doping results in $E_F$-dependent blue shift and narrowing of the G~peak\cite{pisana,Yan07,das1,das2} according to\cite{Ando06,pisana,lazzeriM}:
\begin{widetext}
\begin{eqnarray}
&&\hbar\Delta\mathrm{Pos}(\mathrm{G})_{E_F}=\frac{\lambda_\Gamma}{2\pi}\left[|E_F|+\frac{\hbar\mathrm{Pos}(\mathrm{G})_0}{4}\ln\left|\frac{2E_F-\hbar\mathrm{Pos}(\mathrm{G})_0}{2E_F+\hbar\mathrm{Pos}(\mathrm{G})_0}\right|\right],\\
&&\mathrm{FWHM}(\mathrm{G})_{E_F}=FWHM(\mathrm{G})_0\left\{f_F[-\hbar\mathrm{Pos}(\mathrm{G})_0/2-E_F]-f_F[\hbar\mathrm{Pos}(\mathrm{G})_0/2-E_F]\right\}\nonumber\\
\end{eqnarray}
\end{widetext}
where $f_F(E)$ is the Fermi-Dirac distribution at energy~$E$, $\mathrm{Pos}(\mathrm{G})_0$ and $\mathrm{FWHM}(\mathrm{G})_0=\lambda_\Gamma\mathrm{Pos}(\mathrm{G})_0/4$ are the G peak position and width for zero doping, $\lambda_\Gamma$ is the dimensionless EPC for the LO phonons at $\bf\Gamma$. The logarithmic singularity in SLG can be washed out by disorder. Nevertheless it was observed in both SLG\cite{pisana} and, far more clearly, in BLG\cite{Yan08,das2}. Note that $\mathrm{FWHM}(\mathrm{G})_0$ and $\mathrm{Pos}(\mathrm{G})_0$ can give an accurate measure of $\lambda_\Gamma\sim0.03$\cite{piscanec,BPF09}. Measurement of $\mathrm{Pos}(\mathrm{G})$ combined with FWHM(G) can then be used to estimate doping of an arbitrary sample, e.~g., due to charged impurities\cite{cinzapl}.
\begin{figure*}
\centerline{\includegraphics [width=170mm]{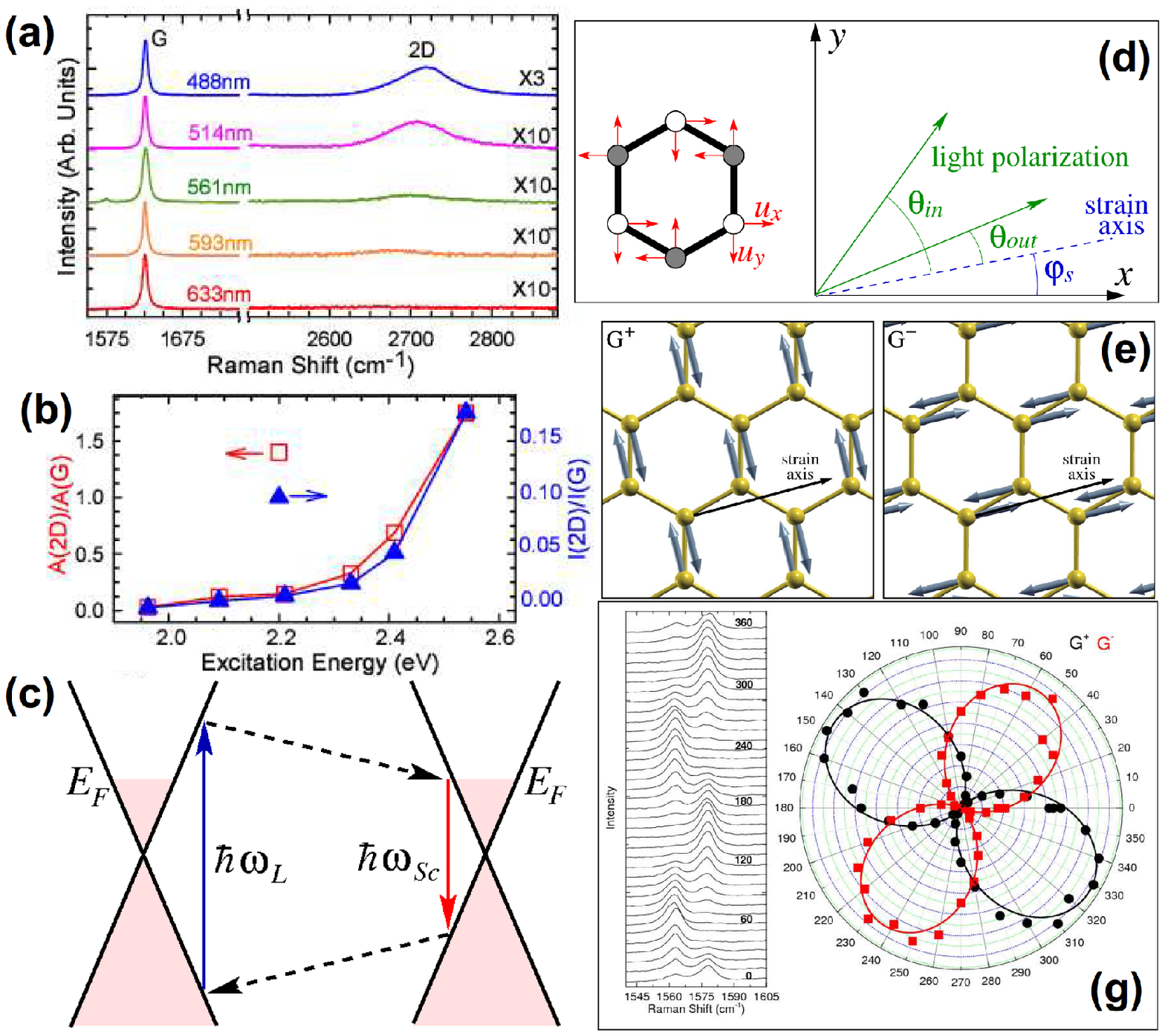}}
\caption{\textbf{Doping and strain.} \textbf{(a)}~Raman spectra of highly doped SLG measured at 488, 514 532, 561, 593 and 633~nm, normalized to have the same G band intensity\cite{tandoping}. \textbf{(b)}~$A(2\mathrm{D})/A(\mathrm{G})$ and $I(2\mathrm{D})/I(\mathrm{G})$ as a function of excitation energy\cite{tandoping}. \textbf{(c)}~Schematic diagram of SLG band structure and 2D Raman processes in doped graphene (analogous to Fig.~2 of the main text). \textbf{(d)}~Geometry for strain measurements. \textbf{(e)}~G~peak splitting by strain. \textbf{(f)}~Sample orientation determined by polarized measurement under uniaxial strain.}
\label{fig:spectrum}
\end{figure*}

The above results apply to relatively weak doping, $|E_F|$ small compared to $\hbar\omega_L/2$. In the past few years, much higher doping levels have been achieved\cite{Kalbac2010,Chen2011,tandoping}. One of the effects of high doping is on the G peak intensity, as can be understood from Fig.~2a of the main text. Doping changes occupations of electronic state and, since transitions from an empty state or to a filled state are impossible, it can effectively exclude some regions of~$\vec{k}$ from contribution to the Raman matrix element. Due to suppression of destructive interference, this leads to an enhancement of the G peak when $|E_F|$ matches $\hbar\omega_L/2$, as predicted theoretically\cite{NJP} and observed experimentally\cite{Kalbac2010,Chen2011}.

The most dramatic effect of high doping is on the 2D peak, which is suppressed when the conduction band becomes filled at the energy probed by the laser\cite{Kalbac2010,Chen2011,tandoping}. Fig.\ref{fig:spectrum}(a) plots the Raman spectra measured at 488, 514, 561, 593 and 633nm for a highly doped graphene sample\cite{tandoping}. $A(2\mathrm{D})/A(\mathrm{G})$ and $I(2\mathrm{D})/I(\mathrm{G})$ are plotted as a function of excitation energy in Fig.\ref{fig:spectrum}(b). The trend of these intensity ratios can be understood from Fig.\ref{fig:spectrum}(c). The frequencies of the absorbed and emitted photons $\omega_L,\omega_{Sc}$ differ by $\mbox{Pos}(2\mathrm{D})$: $\omega_{Sc}=\omega_L-\mbox{Pos}(2\mathrm{D})$. There are three cases: (i) when $\omega_L,\omega_{Sc}>2|E_F|/\hbar$, all processes are allowed and the 2D band is observed, (ii) when $\omega_{Sc}<2|E_F|/\hbar<\omega_L$, the photon absorption is allowed but the phonon emission is excluded by Pauli blocking; (iii) when $\omega_L,\omega_{Sc}<2|E_F|/\hbar$, both photon absorption and phonon emission are blocked. Therefore, only when $2|E_F|/\hbar<\omega_L-\mbox{Pos}(2\mathrm{D})$, the 2D band is observable. Thus, the absence of the 2D band in the Raman spectra in Fig.~\ref{fig:spectrum}(a) at 1.96eV (633nm) indicates that their $E_F$ should be larger than 0.81eV. From the sharp increase in $A(2\mathrm{D})/A(\mathrm{G})$ and $I(2\mathrm{D})/I(\mathrm{G})$ when $\hbar\omega_L$ moves from 2.21eV (561nm) to 2.09eV (593nm), one can deduce that $\hbar\omega_S$ corresponding to $\hbar\omega_L$=2.09$\:\mbox{eV}$ is close to $2E_F$. A similar effect is expected on other DR and TR Raman peaks, such as D, D$'$, 2D$'$, D+D$''$, etc.

At sufficiently high doping, $E_F>E_L/2-\hbar\Omega_q^{TO}$, an additional decay channel opens for the finite-$q$ TO phonon: production of intraband electron-hole pairs\cite{Sasaki2012}, which results in an additional contribution to the width of the 2D band, as observed in Ref.~\onlinecite{Chen2011}.
\section{Magnetic field}
Several effects of the perpendicular magnetic field on Raman spectra of graphene have been observed in recent years. When a perpendicular magnetic field is applied, the electronic trajectories are no longer straight, but circular. This modifies the backscattering condition, so the emitted phonons have smaller momenta than those given by Eq.~(2) of the main text. This results in a red shift and additional 2D peak broadening\cite{Faugeras2010}.

The effect of coupling between $\pi$ electrons and $\bf{\Gamma}$ phonons assumes a peculiar form when a sufficiently strong ($10-30$~T) perpendicular magnetic field $B$ is applied\cite{Ando07,falko}. The correction to the phonon frequency depends on the nature of the electronic states, and in a magnetic field the latter are quantized into discrete Landau levels with energies $E_n(B)$: $E_n(B)=(\mathop{\mathrm{sign}}{n})\sqrt{2|n|v_F^2eB/c}$, $n=\ldots,-1,0,1,2,\ldots$. The $\bf{\Gamma}$ phonons couple to electronic transitions $n\to{n'}$ between Landau levels $n,n'$, which satisfy the selection rule $|n|-|n'|=\pm{1}$, thus having frequencies: $\hbar\Omega_m(B)=E_m(B)+E_{m+1}(B)$ for the undoped graphene. At certain $B=B_m$ the resonance condition $\Omega_m(B)=\Omega_\vec{\Gamma}$ is satisfied and, due to electron-phonon interaction, the phonon becomes strongly coupled to the electronic transition\cite{Ando07,falko}. This coupling manifests in a series of avoided crossings in Pos(G) as a function of $B$ near $B=B_m$\cite{Faugeras2009,pellegrini,Faugeras2011,smi11,smi12}, whose strength depends on the filling factor, as observed in Ref.~\onlinecite{smi12}. Besides providing another EPC estimate, magneto-Raman can also probe the electronic Landau levels, from which $v_F$ can be extracted. An analogous effect was predicted for BLG\cite{Ando2007BLG}, and observed for four-layer graphene\cite{Faugeras2012}. Since the magneto-phonon resonance is strongly sensitive to the electronic band structure, its Raman spectrum is quite different for different numbers of layers, and thus can be used for its determination.

Raman spectroscopy can also probe scattering of photons by electronic excitations. In pristine graphene electronic excitations have a continuous structureless spectrum\cite{wallace}, which does not lead to any sharp features. However, in a strong magnetic field, when the electronic spectrum consists of discrete Landau levels, the electronic inter-Landau-level excitations give rise to sharp $B$-dependent peaks in the Raman spectrum (instead of a phonon, an electron-hole pair is emitted)~\cite{Kashuba2009,Mucha,Kashuba2012}. The selection rule for the Raman-active electronic transitions $n\to{n}'$ is $|n'|=|n|$ when excitation and detection have the same circular polarization, and $|n'|=|n|\pm{2}$ for different circular polarizations. The Raman peaks are observed at frequencies $\hbar\Omega_m=2E_m$ and $\hbar\Omega_m=E_m+E_{m+2}$, respectively\cite{Faugeras2011,Kossacki2011,smi11}
\section{Uniaxial and Biaxial Strain}
Strain arises when a crystal is compressed or stretched out of equilibrium. The stiffness tensor provides the constitutive relation between applied stress and strain. Tensile strain usually gives phonon softening, and the opposite for compressive strain. The rate of these changes is summarized in the Gr\"uneisen parameters, which also determine the thermomechanical properties\cite{grimvall}. Refs.~\onlinecite{Mohi,Hone} subjected graphene to uniaxial strain. The doubly degenerate $E_{2g}$ optical mode was shown to split in two components, one along the strain and the other perpendicular (Fig.~\ref{fig:spectrum}e). This leads to G~peak splitting into two bands\cite{Mohi,Hone}, named G$^+$ and G$^-$ by analogy with the effect of curvature on the nanotube G peak\cite{piscaprb}. Both red-shift with increasing strain, and their splitting increases\cite{Mohi}.

The Gr\"uneisen parameter for the $E_{2g}$ phonon, $\gamma_{E_{2g}}$, is defined as \cite{grimvall} $\gamma_{E_{2g}}=-(1/\Omega^{E_{2g}})(\partial\Delta\Omega^{E_{2g}}_h/\partial\varepsilon_{h})$ where $\varepsilon_{h}=\varepsilon_{ll}+\varepsilon_{tt}$ is the hydrostatic component of the applied uniaxial strain, $l$ is the longitudinal direction, parallel to the strain, and $t$ is the direction transverse to it; $\Omega^{E_{2g}}$ is $\mathrm{Pos}(\mathrm{G})$ at zero strain, and $\Delta\Omega^{E_{2g}}_h$ is the shift resulting from the hydrostatic component. The hydrostatic deformation preserves the symmetry of the crystal, so it does not lead to splitting. The shear deformation potential $\beta_{E_{2g}}$ is defined analogously\cite{Thomsen,Mohi}:$\beta_{E_{2g}}=(1/\Omega^{E_{2g}})(\partial\Delta\Omega^{E_{2g}}/{\partial\varepsilon_{s}})$, where $\varepsilon_{s}=\varepsilon_{ll}-\varepsilon_{tt}$ is the strain shear component, and $\Delta\Omega_s^{E_{2g}}$ is the splitting due to shear. Under uniaxial strain the shifts of the G$^\pm$ peaks are\cite{Sakata,Thomsen,Mohi}:
\begin{equation}\begin{split}
\Delta\mathrm{Pos}(\mathrm{G}^\pm){}&{}=
\Delta\Omega_h^{E_{2g}}\pm\frac{\Delta\Omega_s^{E_{2g}}}{2}=\\
&{}={}-\gamma_{E_{2g}}\Omega^{E_{2g}}\left(\varepsilon_{ll}+\varepsilon_{tt}\right)\pm
\beta_{E_{2g}}\Omega^{E_{2g}}\,\frac{\varepsilon_{ll}-\varepsilon_{tt}}2.
\end{split}\end{equation}
Ref.~\onlinecite{Mohi} obtained $\partial\mathrm{Pos}(\mathrm{G}^+)/\partial\varepsilon=-10.8\:\mbox{cm}^{-1}/\%$, $\partial\mathrm{Pos}({\mathrm{G}^-})/\partial\varepsilon=-31.7\:\mbox{cm}^{-1}/\%$, $\partial\mathrm{Pos}(2\mathrm{D})/\partial\varepsilon=-64\:\mbox{cm}^{-1}/\%$, $\partial\mathrm{Pos}(2\mathrm{D}')/\partial\varepsilon\sim=-35\:\mbox{cm}^{-1}/\%$, $\gamma_{E_{2g}}=1.99$, $\beta_{E_{2g}}=0.99$.

The phonon displacements corresponding to the G$^\pm$ peaks are orthogonal \cite{Sakata,Thomsen,Mohi}: $E^+_{2g}$ is perpendicular to the strain (thus experiencing smaller softening), and $E^-_{2g}$ parallel (Fig.~\ref{fig:spectrum}e). With both polarizer and analyzer, the corresponding polarization vectors $\mathbf{e}_{in},\mathbf{e}_{out}$ have definite orientations: $e^x_{in,out}=\cos(\theta_{in,out}+\varphi_s)$, $e^y_{in,out}=\sin(\theta_{in,out}+\varphi_s)$, where $\theta_{in,out}$ is the (known) angle between $\mathbf{e}_{in,out}$ and the strain axis, and $\varphi_s$ is the (unknown) angle between the strain axis and the crystallographic axis. The matrix elements corresponding to emission of longitudinal and transverse phonons are proportional to $-\sin(\theta_{in}+\theta_{out}+3\varphi_s)$ and $\cos(\theta_{in}+\theta_{out}+3\varphi_s)$. The intensities are given by their squares:
\begin{equation}\begin{split}\label{IlIt=}
&I(\mathrm{G}^-)\propto\sin^2(\theta_{in}+\theta_{out}+3\varphi_s),\\ &I(\mathrm{G}^+)\propto\cos^2(\theta_{in}+\theta_{out}+3\varphi_s).
\end{split}\end{equation}
These allow to determine the sample orientation with respect to strain (Fig.~S3f).

It is important to consider that, while G probes the same centre-zone phonon under strain, this is not necessarily the case for 2D, 2D$'$ (Ref.\onlinecite{Mohi}). Any change in the band structure during strain will vary the actual phonons probed, as well as modifying the phonon frequencies. Thus, the relationship between phonon Gruneisen parameters and 2D, 2D$'$ variation is in principle more complex than for G. Indeed, while biaxial strain does not move the relative positions of the Dirac cones, uniaxial strain changes them\cite{castroneto}. This can have a significant influence in double-resonant and triple-resonant processes. While 2D$'$ is intra-valley, 2D requires scattering from one cone to the other. Thus, its wave vector is determined by the relative distance of the Dirac cones and by the laser energy. For a fixed excitation energy, one measures the combination of the 2D shift due to strain, and a possible additional shift due to the fact that the relative movement of the Dirac cones changes phonon wavevector. For an asymmetric movement this can lead to peak broadening\cite{Mohi}, and splitting\cite{heinzinner,soninner}. The 2D broadening and splitting can give significant information on the nature of the triple-resonant process and the phonons involved, as well as on the band structure under strain, and the orientation of the strain axis with respect to the C$-$C bond\cite{Mohi,Hone,heinzinner,soninner}. For 2D$'$ the relative cone movement has no consequence. However, for 2D, 2D$'$ other effects could be induced by Fermi and phonon group velocity renormalisation with strain.

Note that for biaxial strain at least the effects due to relative movement of Dirac cones are absent. Then, Raman experiments under biaxial strain are suited to measure the D Gruneisen parameter. Ref.~\onlinecite{halsall} performed these, and found no G or 2D peak splitting, as well as shifts and Gruneisen parameters in agreement with those deduced from the uniaxial measurements in Ref.~\onlinecite{Mohi}.
\section{Temperature}
Due to anharmonic effects and phonon-phonon interactions, peak positions and FWHM depend on temperature~$T$\cite{Cowley}. The investigation of the Raman peaks a function of $T$ can provide valuable information on the anharmonic terms in potential energy and EPC\cite{Bonini2007,pisana}. In graphene $\mbox{FWHM}(\mathrm{G})$ depends on $T$ in a peculiar way, slightly decreasing for low $T$, then increasing for $T>700$~K. This can be written as\cite{Bonini2007,pisana}: $\mbox{FWHM}(\mathrm{G})_T=\mbox{FWHM}(\mathrm{G})_T^{EPC}+\mbox{FWHM}(\mathrm{G})_T^{an}$ where $\mbox{FWHM}(\mathrm{G})^{an}_T$ is the anharmonic contribution due to interaction with other phonons, and $\mbox{FWHM}(\mathrm{G})^{EPC}$ the interaction with \textit{e-h} pairs, given in Eq.~(S9), while $\mathrm{FWHM}(\mathrm{G})^{an}$ was derived numerically\cite{Bonini2007}. Eq.~(S9) shows that the $E_{2g}$ phonons damping due to decay in \textit{e-h} pairs decreases with $T$ due to partial Pauli blocking\cite{pisana}. Approximate expression for $\mbox{FWHM}(\mathrm{G})^{an}$ can be derived assuming that phonons decay in 2 (3-phonon scattering) or 3 phonons (4-phonon scattering) with the same energy\cite{Balkanski}. In this case, $\mathrm{FWHM}(\mathrm{G})_T^{an}=A\left[1+\frac{2}{e^x-1}\right]+B\left[1+\frac{3}{e^y-1}+\frac{3}{(e^y-1)^2}\right]$ where $x$=$\hbar Pos(\mathrm{G})_0/2k_BT$ and $y=\hbar \mbox{Pos}(\mathrm{G})_0/(3k_BT)$, and $A,B$ are constants.

The temperature dependence of $\mbox{Pos}(\mathrm{G})$ has a quasi-harmonic term, from thermal expansion, and an anharmonic one accounting for phonon-phonon scattering. Ref.\onlinecite{Balkanski} proposed a simple model for the anharmonic phonon-phonon coupling including 3-phonon (3-ph) and 4-phonon (4-ph) processes. Then $\mbox{Pos}(\mathrm{G})_T=\mbox{Pos}(\mathrm{G})_0+\Delta\mbox{Pos}(\mathrm{G})_T$, where $\Delta\mbox{Pos}(\mathrm{G})_T=C\left[1+\frac{2}{e^x-1}\right]+D\left[1+\frac{3}{e^y-1}+\frac{3}{(e^y-1)^2}\right]$, where $C,D$ are constants, and $x,y$ are the same as above.
\section{Isotopic composition}
$^{13}$C can be used to label graphene as well as nanotubes in order to measure fundamental properties or, for example, to reveal growth processes\cite{cai}. The peak position of an isotope enriched sample is $\omega_{C_{13}}=\omega_{C_{12}}\cdot\sqrt{m_{12}/m_{13}}$ where $\omega_{C_{12}}$ and $\omega_{C_{13}}$ are the frequencies for full $^{12}$C and $^{13}$C samples, $m_{12}$ and $m_{13}$ the atomic masses of $^{12}$C and $^{13}$C. The isotopic shift is thus:
\begin{equation}
\frac{\partial\omega}{\partial C_{13}\%}=\frac{1}{100}\left(\omega^{0}_{C_{12}}\cdot\sqrt{\frac{m_{12}}{m_{13}}}-\omega^{0}_{C_{12}}\right)\approx- \frac{0.0392}{100}\omega^{0}_{C_{12}}
\end{equation}
Isotopic disorder also widens FWHM(G), with a maximum around $50\%$ $^{13}$C.

\section{Raman and graphene properties}
\subsection{Raman and electrical transport} Raman spectroscopy is ideal to probe defects, electron-phonon and e-e interactions. Thus it can link sample quality to mobility\cite{fuhrer,ni}, or EPC with current saturation\cite{berciaud,freitag}.

\subsection{Raman and heat transport} The $\mbox{Pos}(G)$ temperature coefficient enabled the determination of thermal conductivity in SLG and graphene layers\cite{balandin,balandin2}.

\subsection{Detection of heteroatoms} Functionalization and doping are an ever growing field. Any covalent bond gives a D~peak. This is an \textit{indirect} signature of C--H, C--O, C--F, C--N, C--Si, \textit{etc.}, bonding\cite{elias,gokus,Ferrari00}, when visible Raman is performed, since the C--C $sp^2$ bonds always prevail\cite{Ferrari00}. On the other hand UV Raman spectroscopy can directly probe the heteroatomic vibrations\cite{Ferrari00,prbcn}, and as such could be extremely useful, especially for wide band gap compounds derived from graphene, and to probe SiC grown samples. Extreme care is needed with deep UV Raman since this can easily damage the sample and break the very bonds one wishes to study. UV Raman also allows to probe C--C $sp^3$ vibrations, otherwise overshadowed for visible excitation\cite{Ferrari00}.

\section*{Acknowledgements}
We thank E. Lidorikis, S. Piscanec, P.H. Tan, S. Milana, D. Yoon, A. Lombardo, A. Bonetti, C. Casiraghi, F. Bonaccorso, G. Savini, N. Bonini, N. Marzari, T. Kulmala, A. Jorio, M. A. Pimenta, G. Cancado, R. Ruoff, R. A. Nair, K. A. Novoselov, L. Novotny, A. K. Geim, C. Faugeras, M. Potemski, and I. L. Aleiner for useful discussions. ACF acknowledges funding from the Royal Society, the European Research Council Grant NANOPOTS, EU Grants RODIN, MEM4WIN, and CareRAMM, EPSRC grants EP/K01711X/1, EP/K017144/1, EP/G042357/1 and Nokia Research Centre, Cambridge.

\end{document}